\documentclass[12pt,preprint2]{aastex6}
\usepackage{graphicx}
\usepackage{natbib}
\usepackage{amsmath}
\usepackage{tgcursor}
\usepackage{booktabs}

\setcitestyle{notesep={ }}

\begin{document}

\title{A survey of CH$_3$CN and HC$_3$N in protoplanetary disks}

\author{Jennifer B. Bergner\altaffilmark{1}, Viviana G. Guzm\'an\altaffilmark{2}, Karin I. \"Oberg\altaffilmark{3}, Ryan A. Loomis\altaffilmark{3}, Jamila Pegues\altaffilmark{3}}

\altaffiltext{1}{Harvard University Department of Chemistry and Chemical Biology, Cambridge, MA 02138, USA}
\altaffiltext{2}{Joint ALMA Observatory (JAO), Alonso de C\`ordova 3107, Vitacura, Santiago, Chile}
\altaffiltext{3}{Harvard-Smithsonian Center for Astrophysics, Cambridge, MA 02138, USA}
 
\begin{abstract}
The organic content of protoplanetary disks sets the initial compositions of planets and comets, thereby influencing subsequent chemistry that is possible in nascent planetary systems.  We present observations of the complex nitrile-bearing species CH$_3$CN and HC$_3$N towards the disks around the T Tauri stars AS 209, IM Lup, LkCa 15, and V4046 Sgr as well as the Herbig Ae stars MWC 480 and HD 163296.  HC$_3$N is detected towards all disks except IM Lup, and CH$_3$CN is detected towards V4046 Sgr, MWC 480, and HD 163296.  Rotational temperatures derived for disks with multiple detected lines range from 29--73K, indicating emission from the temperate molecular layer of the disk.  V4046 Sgr and MWC 480 radial abundance profiles are constrained using a parametric model; the gas-phase CH$_3$CN and HC$_3$N abundances with respect to HCN are a few to tens of percent in the inner 100 AU of the disk, signifying a rich nitrile chemistry at planet- and comet-forming disk radii.  We find consistent relative abundances of CH$_3$CN, HC$_3$N, and HCN between our disk sample, protostellar envelopes, and solar system comets; this is suggestive of a robust nitrile chemistry with similar outcomes under a wide range of physical conditions.
\end{abstract}


\section{Introduction}
Planets form by accreting gas and dust within the protoplanetary disk; the material present in the disk therefore sets the initial composition of planets.  Observations of solar system comets suggest that the solar nebula was rich in volatile organic molecules around the time the comets were formed, usually with abundances of a few percent with respect to water ice \citep[e.g.][]{Mumma2011}.  Understanding how this early inventory of organic molecules developed into the vast complexity of biochemistry is key to the study of the origins of life.  Recent successes in prebiotic syntheses of RNA and protein precursors suggest that nitrile-bearing molecules, characterized by the C$\equiv$N functionality, played a crucial role in prebiotic chemistry \citep{Powner2009, Ritson2012, Sutherland2016}.  From cometary studies, HCN, CH$_3$CN, and HC$_3$N indeed appear to have been common in the young Solar system \citep[][]{Mumma2011,Cordiner2014,LeRoy2015}.  To evaluate whether the nitrile chemistry of the solar nebula is typical, and in turn any implications for the chemical habitability of other planetary systems, observations of planet-forming disks are essential.  

The simple nitrile species CN and HCN were first detected towards protoplanetary disks two decades ago \citep{Dutrey1997, Kastner1997} and have since been observed towards many additional disks \citep[see][]{Dutrey2014}.  Until the advent of ALMA, observational challenges limited our ability to detect and characterize more complex nitrile species in disks.  The first disk detections of HC$_3$N were made by \citet{Chapillon2012} with the IRAM 30m telescope towards GO Tau, LkCa 15, and MWC 480.  CH$_3$CN was first detected in a disk by \citet{Oberg2015} towards MWC 480 using ALMA.  The molecular emission was spatially resolved, allowing radial abundance profiles of CH$_3$CN and HC$_3$N to be derived.  At comet-forming disk radii the abundances were found to be similar to those measured in solar system comets, suggesting that the solar system is not unique in its nitrile chemistry.  

In addition to their relevance to pre-biotic chemistry, CH$_3$CN and HC$_3$N are, along with CH$_3$OH \citep{Walsh2016}, the only large organic molecules detected in protoplanetary disks to date.  Because of this, these molecules are key to furthering our understanding of the growth of organic complexity in disks.  Their abundances and distributions can be used to benchmark astrochemical models of disks, and in turn to help predict the chemistry of complex molecules which cannot currently be directly observed.  This is of particular importance for gaining insights into the ice compositions in disks, which can be constrained only through chemical models.

To date, the number of disks with well-characterized nitrile abundances is small, and it is unclear (i) whether other disks commonly host similar nitrile abundances as the solar nebula, and (ii) how robust the nitrile chemistry is across different circumstellar environments.  A larger sample of observations is required to obtain constraints on the nitrile chemistry in disks.

Here, we present observations of the complex nitrile molecules HC$_3$N and CH$_3$CN towards a diverse sample of six protoplanetary disks: our targets span over an order of magnitude in luminosity and disk age and represent both transition disks and full disks.  In Section \ref{sec_methods} we describe the observations and data reduction.  The observational results are presented in Section \ref{sec_obsresults}.  In Section \ref{sec_model} a parametric model is used to obtain abundance profiles of CH$_3$CN an HC$_3$N towards the bright sources MWC 480 and V4046 Sgr.  Finally, in Section \ref{sec_discussion} we comment on implications for the nitrile chemistry in other circumstellar systems based on our findings.

\section{Methods}
\label{sec_methods}
\subsection{Observations}
During ALMA Cycle 2, the HC$_3$N 27--26 transition as well as the CH$_3$CN 14--13 K-ladder were observed for the disks AS 209, HD 163296, IM Lup, LkCa 15, MWC 480, and V4046 Sgr (project code 2013.1.00226).    Disk and stellar properties for each source are listed in Table 1.  The MWC 480 data were previously presented in \citet{Oberg2015}; these lines were re-analyzed in this work to ensure that all sources were treated consistently.  During Cycles 3 and 4, as part of a line survey in disks (project code 2013.1.01070.S), the HC$_3$N 31--30 and 32--31 transitions and the CH$_3$CN 15--14 and 16--15 K-ladders were observed towards MWC 480 and LkCa 15.

A detailed description of the Cycle 2 observations can be found in \citet{Huang2017}.  Briefly, from 2014 to 2015 Band 6 observations were taken at two spectral settings, 1.1mm and 1.4mm, containing 14 and 13 narrow spectral windows, respectively.  Baselines spanned 18 to 650m, and the total on-source time was $\sim$20 minutes per source.  Amplitude and phase calibration, as well as frequency bandpass calibration, were performed using observations of a quasar.  Flux calibration was performed using observations of either Titan or a quasar.  A portion of the CH$_3$CN 14--13 ladder (14$_0$--13$_0$, 14$_1$--13$_1$, and 14$_2$--13$_2$) is contained in a spectral window of 59 MHz with a channel width of 61 kHz (0.071 km/s) in the 1.1mm spectral setting.  The HC$_3$N 27--26 transition is contained in a spectral window of 117 MHz, also with a channel width of 61 kHz (0.075 km/s) in the 1.1mm spectral setting.  

\begin{deluxetable*}{lcccccccccc} 
	\tabletypesize{\footnotesize}
	\tablecaption{Disk and star properties of observed sources}
	\tablecolumns{11} 
	\tablewidth{\textwidth} 
	\tablehead{
		\colhead{Source}                            							&
		\colhead{Distance}										&
		\colhead{Spectral}                 			          				&
		\colhead{Age}             		         			 			&
		\colhead{$M_\star$}										&
       		 \colhead{$L_\star$}										&
        		\colhead{$\dot{M}_\star$}									&
		\colhead{Disk Inc.}                          							&
 	        \colhead{Disk PA} 										&
		\colhead{$M_{disk}$} 									&
		\colhead{$v_{\mathrm{LSR}}$}					  			\\
		\colhead{}                            								&
		\colhead{(pc)}											&
		\colhead{type}                 			   		       				&
		\colhead{(Myr)}            		         			 			&
		\colhead{($M_\odot$)}									&
         	\colhead{($L_\odot$)}									&
        		\colhead{(10$^{-9} M_\odot$ yr$^{-1}$)}						&
		\colhead{(deg)}                          								&
        		\colhead{(deg)} 										&
		\colhead{($M_\odot$)} 									&
		\colhead{(km s$^{-1}$)}						  			}		
\startdata
AS 209 & 143$^a$ & K5 & 1.6 & 0.9 & 1.5 & 51 & 38 & 86 & 0.015 & 4.6 \\ 
IM Lup & 161$^a$ & M0 & 1 & 1.0 & 0.93 & 0.01 & 50 & 144.5 & 0.17 & 4.4 \\
LkCa 15 & 140 & K3 & 3-5 & 0.97 & 0.74 & 1.3 & 52 & 60 & 0.05-0.1 & 6.3 \\
V4046 Sgr & 72 & K5, K7 & 24 & 1.75 & 0.49, 0.33 & 0.5 & 33.5 & 76 & 0.028 & 2.9 \\
MWC 480 & 142$^a$ & A4 & 7 & 1.65 & 11.5 & 126 & 37 & 148 & 0.11 & 5.1 \\
HD 163296 & 122 & A1 & 4 & 2.25 & 30 & 69 & 48.5 & 132 & 0.17 & 5.8 \\
\enddata
\tablecomments{Table adapted from \citet{Huang2017}, where a full list of references can be found.}
\tablenotetext{a}{From Gaia Data Release 1 \citep{Gaia2016}}
\label{table_diskstats}
\end{deluxetable*}

The Cycle 3/4 observations are described in full in \citet{Loomis2017a}.  This work made use of data taken in two Band 7 correlator setups at 0.9mm and 1.1mm, each containing 4 spectral windows of 1920 channels with channel widths of 975 kHz ($\sim$0.99 -- 1.06 km/s for the lines of interest).  Baselines spanned 15 to 650m, and the total on-source integration time was $\sim$20 minutes per source.  Phase calibration, bandpass calibration, and flux calibration were all performed using quasar observations.  

\subsection{Data reduction}
\label{sec_datareduce}
Initial data calibration was performed by ALMA/NAASC staff.  Two rounds of phase self-calibration were performed using the continuum emission from individual spectral windows, except for HD 163296 which was self-calibrated using averaged spectral windows due to weak continuum emission.  Following continuum subtraction, the data cubes were imaged in {\fontfamily{qcr}\selectfont CASA} using the CLEAN task with a 3$\sigma$ noise threshold.  Briggs weighting was used with a robust parameter of 2.0 for all lines except those in the source V4046 Sgr; here, a value of 1.0 was adopted to improve the angular resolution, which was possible due to higher signal-to-noise ratios.  To obtain channel maps, the data were regridded to 0.5 km s$^{-1}$ and 1.1 km s$^{-1}$ spectral resolution for the Cycle 2 and Cycle 3/4 observations, respectively.  CLEAN masks were drawn by hand for lines with obvious emission (V4046 Sgr HC$_3$N 27-26 and CH$_3$CN 14$_0$-13$_0$ and 14$_1$-13$_1$).  For all other lines, the 5$\sigma$ continuum contour was used as the CLEAN mask. 

\begin{figure*}
	\includegraphics[width=\linewidth]{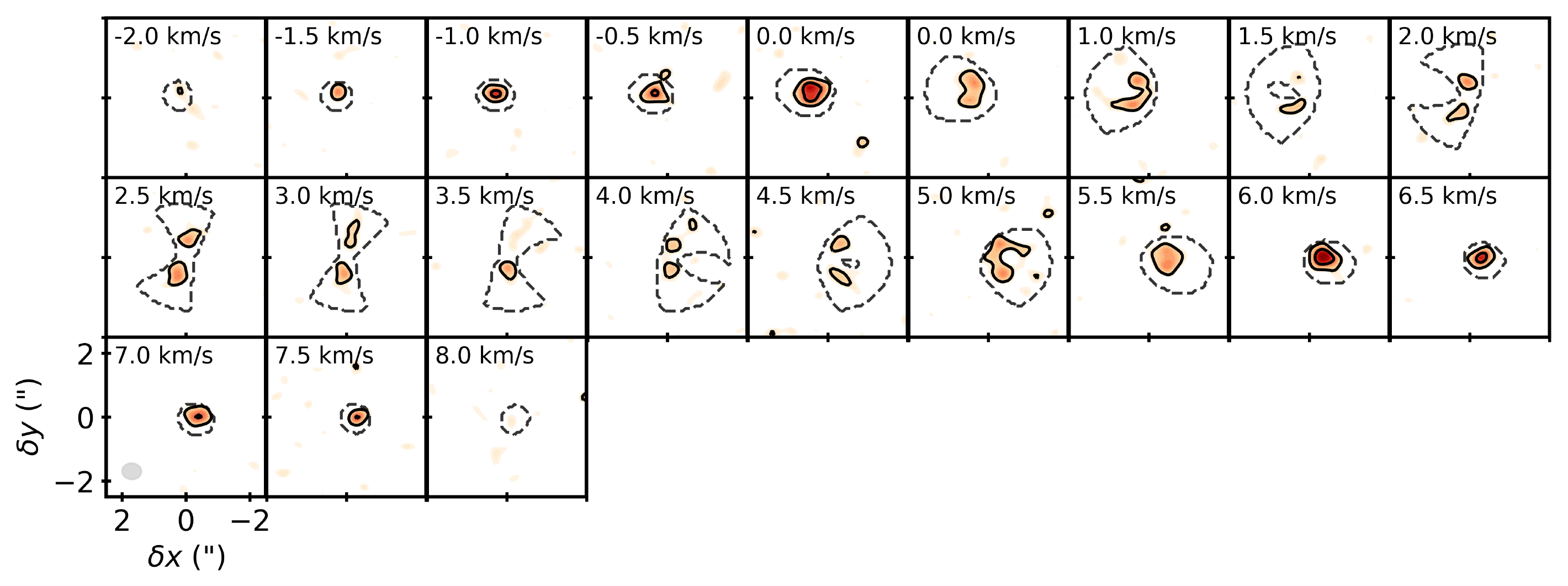}
	\caption{Channel maps with a Keplerian mask overlaid for the HC$_3$N 27-26 transition in V4046 Sgr.  Emission below the 2$\sigma$ level is not shown.  The synthesized beam is shown in the bottom left panel.  Contours correspond to 3, 6, and 10$\times$ rms.}
\label{chan_hc3n_v4046sgr}
\end{figure*}

\begin{deluxetable*}{lllccccc}
	\tabletypesize{\footnotesize}
	\tablecaption{Line observations (detections \& non-detection upper limits)}
	\tablecolumns{8} 
	\tablewidth{\textwidth} 
	\tablehead{
		\colhead{Transition}                           							&
		\colhead{Frequency}                         							&
		\colhead{Source}										&
		\colhead{Beam}                 			          				&
		\colhead{Beam PA}	             					 			&
		\colhead{Channel rms$^{a}$}								&
		\colhead{Mom. zero rms$^{b}$}								&
        		\colhead{Int. Flux Density$^{c}$}									\\
		\colhead{}                            								&
		\colhead{(GHz)}                        								&
		\colhead{}												&
		\colhead{ (") }                 			   		       				&
		\colhead{($^\mathrm{o}$)}            		         			 	&
		\colhead{(mJy beam$^{-1}$)}								&
		\colhead{(mJy beam$^{-1}$ km s$^{-1}$)}						&
        		\colhead{(mJy km s$^{-1}$)}								}		
\startdata
CH$_3$CN 14$_0$--13$_0$ & 257.5274 & AS 209 & 0.51 x 0.45 & -66.2 & 3.1 & 4.3 & $<$39  \\ 
 &  & IM Lup & 0.47 x 0.43 & 83.4 & 3.0 & 4.1 & $<$33  \\ 
 &  & LkCa 15 & 0.67 x 0.51 & -15.5 & 2.8 & 4.7 & $<$28  \\ 
 &  & V4046 Sgr & 0.59 x 0.48 & 84.9  & 2.9 & 7.9 & 299 $\pm$ 94$^{d}$ \\ 
 &  & MWC 480 & 0.76 x 0.49 & -8.6  & 2.8 & 4.8 & 42 $\pm$ 11 \\ 
 &  & HD 163296 & 0.60 x 0.46 & -87.1  & 2.3 & 4.2 & 53 $\pm$ 11 \\ 
 \addlinespace
CH$_3$CN 14$_1$--13$_1$ & 257.5224 & V4046 Sgr & 0.59 x 0.48 & 84.9  & 2.9 & 7.9 & 259 $\pm$ 82$^{d}$ \\ 
\addlinespace
CH$_3$CN 14$_2$--13$_2$ & 257.5076 & V4046 Sgr & 0.59 x 0.48 & 84.9  & 3.0 & 4.7 & 113 $\pm$ 17 \\ 
\addlinespace
CH$_3$CN 15$_0$--14$_0$ & 275.9156 & LkCa 15 & 1.17 x 1.04 & -43.6 & 2.8 & 8.7 & $<$30  \\ 
 &  & MWC 480 & 1.30 x 1.07 & -7.3  & 2.7 & 10.0 & 44 $\pm$ 23$^{d}$ \\ 
 \addlinespace
CH$_3$CN 15$_1$--14$_1$ & 275.9103 & MWC 480 & 1.30 x 1.07 & -7.3  & 2.7 & 10.0 & 42 $\pm$ 19$^{d}$ \\ 
\addlinespace
CH$_3$CN 16$_0$--15$_0$ & 294.3024 & LkCa 15 & 1.35 x 0.85 & -56.5 & 5.3 & 16.0 & $<$54  \\ 
 &  & MWC 480 & 1.41 x 0.86 & -45.6 & 4.8 & 16.0 & $<$67  \\ 
\addlinespace
HC$_3$N 27--26 & 245.6063 & AS 209 & 0.54 x 0.46 & -64.0  & 3.4 & 5.4 & 103 $\pm$ 15 \\ 
 &  & IM Lup & 0.51 x 0.45 & 69.5 & 3.1 & 4.2 & $<$29  \\ 
 &  & LkCa 15 & 0.71 x 0.53 & -14.5  & 2.9 & 4.8 & 44 $\pm$ 9 \\ 
 &  & V4046 Sgr & 0.60 x 0.50 & 86.7  & 3.4 & 6.0 & 352 $\pm$ 20 \\ 
 &  & MWC 480 & 0.81 x 0.51 & -8.5  & 3.1 & 5.3 & 136 $\pm$ 11 \\ 
 &  & HD 163296 & 0.62 x 0.48 & -86.1  & 2.7 & 4.6 & 201 $\pm$ 13 \\ 
\addlinespace
HC$_3$N 31--30 & 281.9768 & LkCa 15 & 1.40 x 0.86 & -56.4 & 4.6 & 12.9 & $<$44  \\ 
 &  & MWC 480 & 1.45 x 0.90 & -44.7  & 4.1 & 11.4 & 92 $\pm$ 15 \\ 
\addlinespace
HC$_3$N 32--31 & 291.0684 & LkCa 15 & 1.40 x 0.86 & -56.4 & 5.0 & 15.2 & $<$48  \\ 
 &  & MWC 480 & 1.38 x 0.90 & -43.7  & 5.0 & 14.2 & 59 $\pm$ 20 \\ 
\enddata
\tablenotetext{a}{For 0.5 km s$^{-1}$ channel widths}
\tablenotetext{b}{Median rms; see Section \ref{sec_datareduce}}
\tablenotetext{c}{3$\sigma$ upper limits are reported for non-detections}
\tablenotetext{d}{Uncertainty includes an additional 30\% of the integrated flux due to line blending}
\label{table_lineobs}
\end{deluxetable*}

A Keplerian mask was applied to the cleaned data cube to obtain moment zero maps, line spectra, and integrated flux densities for each transition.  The use of Keplerian masks as CLEAN templates and for spectral extraction is well established \citep[e.g.][]{Rosenfeld2013b, Loomis2015, Oberg2015}; details on the use of Keplerian masking for moment zero map generation will be presented in \citep{Pegues2017}.  Briefly, to construct appropriate masks, the Keplerian velocity of each image pixel was calculated based on its deprojected radius from the star, assuming disk and stellar parameters taken from the literature (Table 1).  In each velocity channel, only pixels corresponding to the appropriate Keplerian velocity were included, and all other pixels were masked.  The mask outer radii were chosen to encompass all HC$_3$N and CH$_3$CN emission.  The masks were verified to fit the actual disk profile using H$^{13}$CN emission \citep{Guzman2017}, which has more obvious Keplerian structure than HC$_3$N or CH$_3$CN.  An example channel map with its Keplerian mask overlaid is shown in Figure \ref{chan_hc3n_v4046sgr} for the HC$_3$N 27-26 transition in V4046 Sgr; for all other lines and sources, a similar figure can be found in Appendix \ref{app_chanmaps}.  

Since the moment zero maps are produced by summing only emission within the Keplerian mask, each moment zero pixel represents the sum of a different number of channels.  The rms is therefore non-uniform across the moment zero map.  We approximate the moment zero rms by bootstrapping: the same Keplerian mask used to obtain the moment zero map is applied to 1000 off-source positions.  The rms per pixel is determined from the standard deviation of each mask pixel across all off-source moment zero maps.  The median rms value is taken to be the representative moment zero rms, and is quoted in Table 2 and used to draw contours in Figures \ref{fig_mom0spec_main} and \ref{fig_mom0_other}.  The uncertainty in the integrated flux density was estimated from the standard deviation of the integrated fluxes within 1000 off-source Keplerian masks. 

In V4046 Sgr, the CH$_3$CN 14$_0$--13$_0$ line is blended with the 14$_1$--13$_1$ line (spectrum shown in Figure \ref{fig_mom0spec_main}).  Likewise, in MWC 480 the CH$_3$CN 15$_0$--14$_0$ and 15$_1$--14$_1$ lines are blended.  To treat blended lines, two Keplerian masks (one centered on each line) were calculated and used to extract the emission from both lines, and the resulting moment zero map sums the total emission.  To estimate the integrated flux density from each individual transition, we assume that emission is symmetric around the rest velocity of the source.  The integrated flux for the lower-energy transition was therefore assumed to be twice the integrated flux of the lowest-velocity horn (i.e., velocities lower than the source velocity).  The integrated flux of the higher-energy transition was taken to be the integrated flux of the total blended feature minus the integrated flux of the low-energy transition.  To account for the added uncertainties in this procedure an additional error of 30\% the integrated flux was added in quadrature with the bootstrapped integrated flux uncertainty for blended lines.

\section{Observational results}
\label{sec_obsresults}
Based on our observations, we now present the molecular line detections and non-detections in our sample.  For molecules with multiple detections towards a source, we use the rotational diagram method to derive rotational temperatures and column densities.  Finally, we calculate disk-averaged abundance ratios of CH$_3$CN/HCN and HC$_3$N/HCN for all disks by assuming a range of emission temperatures.

\subsection{Molecule detections}
A summary of the line observations is presented in Table 2.  The CH$_3$CN 14$_0$--13$_0$ and HC$_3$N 27--26 transitions were targeted towards all 6 disks and, compared to the other observed CH$_3$CN and HC$_3$N transitions, are the brightest across the sample.  We therefore use these lines to classify whether molecules are detected or not detected in a disk.  A molecule is considered detected if emission $>$3$\times$rms is present within the Keplerian mask in at least three channels.  We also tested a detection criterion of a SNR $>$3 for the integrated flux density, and find the same status of detection vs. non-detection for all transitions.  Based on these criteria,  HC$_3$N is detected towards all disks except IM Lup.  CH$_3$CN is firmly detected towards three of six disks (HD 163296, MWC 480, and V4046 Sgr).  In AS 209 and LkCa 15, CH$_3$CN is not seen at significant levels in individual channel maps but shows suggestive features at around a 3$\sigma$ level in the moment zero maps, as well as positive integrated intensities in the radial profiles.  Higher-sensitivity follow-up observations are required to confirm if these are indeed detections.  For subsequent analysis these lines are treated as non-detections (3$\sigma$ upper limits).

\begin{figure*}
	\includegraphics[width=0.9\textwidth]{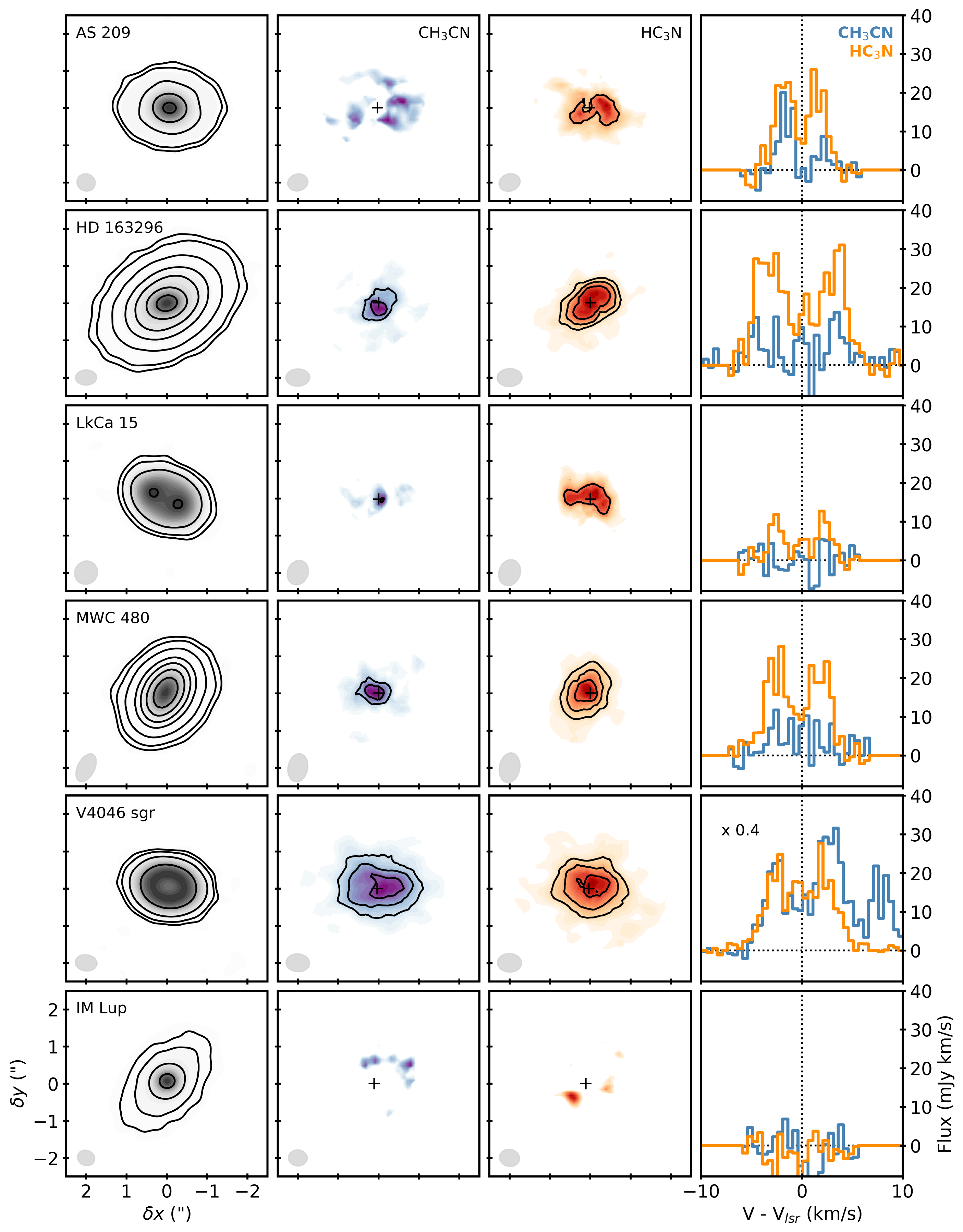}
	\caption{First column: dust continuum maps (1.1mm), with contour lines corresponding to 5, 10, 30, 100, 200, 400, and 800 $\sigma$ emission.  Second and third columns: CH$_3$CN 14$_0$-13$_0$ and HC$_3$N 27-26 moment zero maps, integrated within the Keplerian masks shown in Appendix \ref{app_chanmaps}.  Emission below the minimum rms value is not shown.  Contours correspond to 3, 6, and 10$\times$ the median moment zero rms (Section \ref{sec_datareduce}).  For all moment zero maps, the synthesized beam is shown in the lower left corner, and the continuum centroid is marked with a +.  Fourth column: disk-integrated spectra extracted with Keplerian masks for CH$_3$CN 14$_0$-13$_0$ (blue lines) and HC$_3$N 27-26 (orange lines).  For V4046 Sgr, the CH$_3$CN line is blended with the 14$_1$-13$_1$ transition.}
\label{fig_mom0spec_main}
\end{figure*}

Figure \ref{fig_mom0spec_main} shows the disk continuum maps as well as the CH$_3$CN 14$_0$--13$_0$ and HC$_3$N 27--26 line maps and spectra for each disk.  In all cases the molecular emission is more compact than the continuum emission.  Comparing the nitrile emission across the sample, V4046 Sgr is by far the strongest emitter, with strong detections of both CH$_3$CN and HC$_3$N.  Next strongest are HD 163296 and MWC 480, which both host strong HC$_3$N and moderate CH$_3$CN emission.  AS 209 and LkCa 15 both exhibit moderate HC$_3$N and tentative CH$_3$CN emission, and neither molecule is detected in IM Lup. 

\begin{figure*}
	\includegraphics[width=\linewidth]{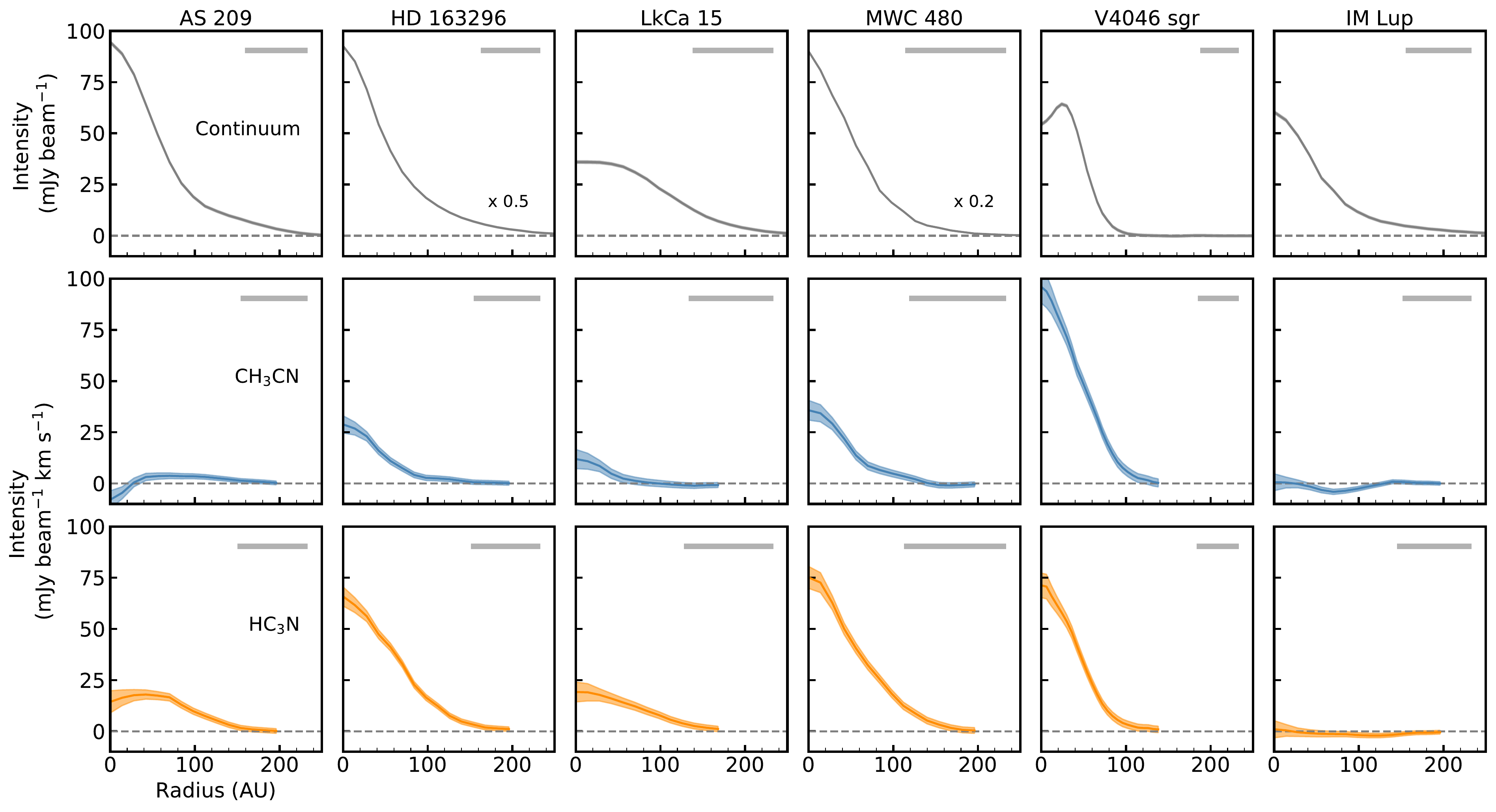}
	\caption{Deprojected and azimuthally averaged radial profiles of the 1.1mm continuum (grey), CH$_3$CN 14$_0$-13$_0$ transition (blue), and HC$_3$N 27-26 transition (orange) in each disk.  Ribbons represent the uncertainty in intensities at each radial distance. Gray bars represent the restoring beam major axis.  The HD 163296 and MWC 480 continuum intensities have been scaled down by the indicated amount for clarity.}
\label{fig_radprof}
\end{figure*}

Figure \ref{fig_radprof} shows the deprojected radial profiles for each transition.  The uncertainties are estimated by dividing the median moment zero rms by the square root of the number of independent measurements (i.e., the number of pixels at each radius divided by the beam size in pixels, to account for beam convolution).  For both CH$_3$CN and HC$_3$N, almost all detections exhibit centrally peaked emission.  The exception is CH$_3$CN and HC$_3$N in AS 209, which peak at larger radii, indicative of a ringed structure; this can also be seen in the moment zero map (Figure \ref{fig_mom0spec_main}).  Although it does not have a large central dust cavity, AS 209 also exhibits a ring-like structure in the molecules H$^{13}$CN,  HC$^{15}$N, DCN, H$^{13}$CO$^+$, and DCO$^+$ \citep{Guzman2017,Huang2017}, possibly due to dust opacity effects.  Additionally, we note that when imaged with a smaller robust factor, HC$_3$N and possibly CH$_3$CN in V4046 Sgr show evidence of a central depression.  Higher-resolution observations are required to confirm the morphology of these molecules at small scales.

\begin{figure*}
	\includegraphics[width=0.9\textwidth]{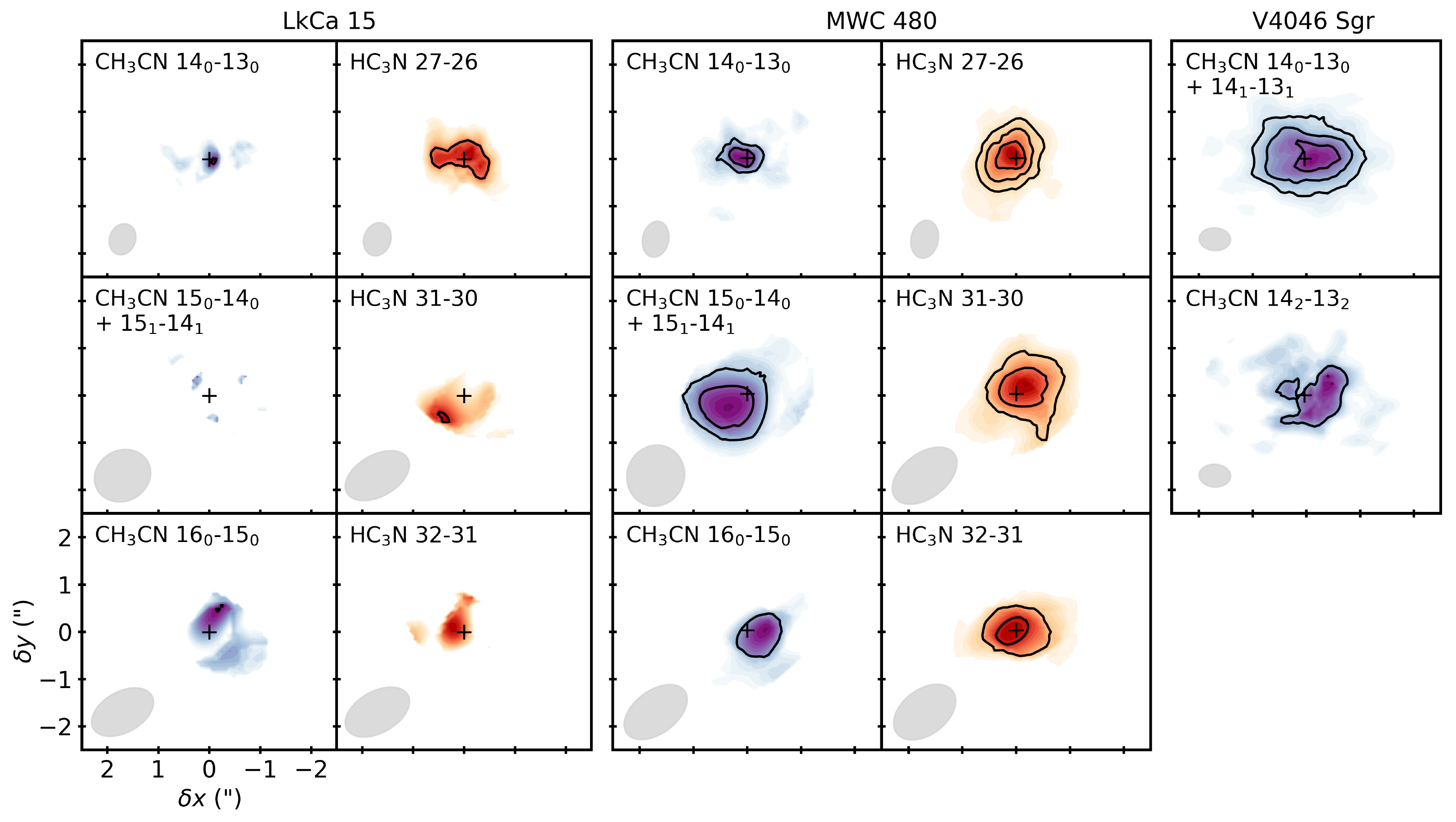}
	\caption{Moment zero maps for molecules with multiple observed transitions, integrated within the Keplerian masks shown in Appendix \ref{app_chanmaps}.  Emission below the minimum rms value is not shown.  Contours correspond to 3, 6, and 10$\times$ rms levels.  For all moment zero maps, the synthesized beam is shown in the lower left corner, and the continuum centroid is marked with a +. }
\label{fig_mom0_other}
\end{figure*}

In V4046 Sgr, MWC 480, and LkCa 15, multiple transitions from the same molecule were observed.  Figure \ref{fig_mom0_other} shows the moment zero maps for these additional transitions.  The CH$_3$CN 14$_1$-13$_1$ and 14$_2$-13$_2$ lines were covered within the same spectral window as the 14$_0$--13$_0$ transition, and in V4046 Sgr  these higher-K lines were strong enough to be detected.  Additionally, for MWC 480 and LkCa 15 the CH$_3$CN 15$_0$-14$_0$ and 16$_0$-15$_0$ and HC$_3$N 31-30 and 32-31 lines were observed in a separate program \citep{Loomis2017a}.  In MWC 480, CH$_3$CN 15$_0$-14$_0$ and HC$_3$N 31-30 and 32-31 were detected.  These additional lines were not detected in LkCa 15.

\subsection{Population diagrams}
\label{sec_multiples}
\begin{figure}
	\includegraphics[width=\linewidth]{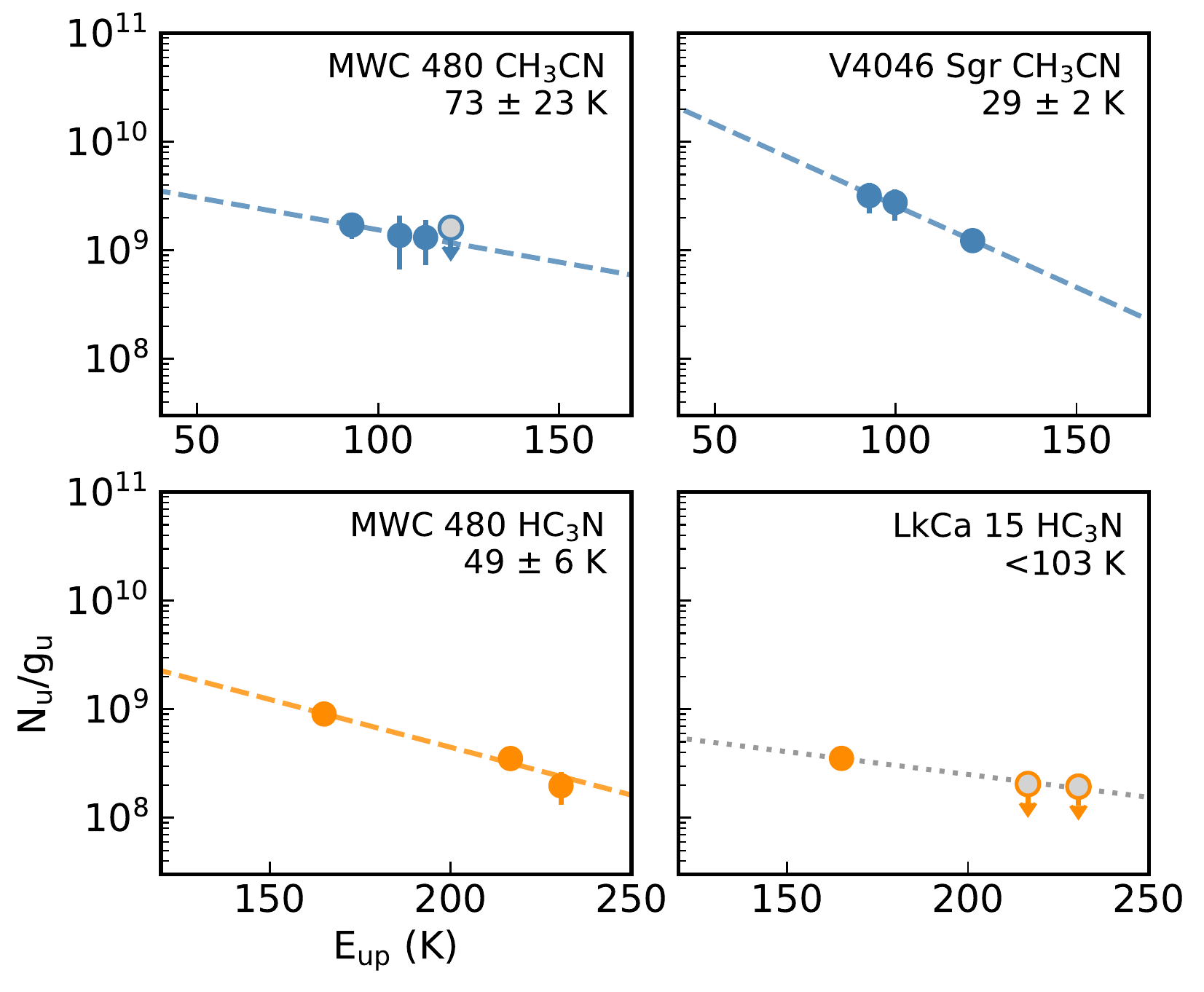}
	\caption{Rotational diagrams for multiple line detections.  Full circles show detections and open circles show 3$\sigma$ upper limits for non-detected lines.  For LkCa 15, the rotational diagram is fit using upper limit constraints since only one line is detected.}
\label{fig_RD}
\end{figure}

For molecules with multiple detections, rotational diagrams can be used to determine the disk-averaged column densities and rotational temperatures of emission (Figure \ref{fig_RD}).  In most cases, only detected lines were used in fitting rotational diagrams; the exception is for HC$_3$N in LkCa 15, as only a single line is detected.  In this case the 3$\sigma$ upper limits on the non-detected transitions were used to obtain an upper limit for the rotational temperature.  

Assuming LTE and optically thin emission, the disk-integrated flux density $S_\nu \Delta V$ can be converted to an upper level population $N_u$ \citep{Goldsmith1999}:

\begin{equation}
N_u = \frac{4\pi S_\nu \Delta V}{A_{ul}\Omega hc},
\end{equation}
where $S_\nu$ is the flux density, $\Delta V$ is the line width, $A_{ul}$ is the Einstein coefficient and $\Omega$ is the solid angle of the source.  For a disk-averaged column density, $\Omega$ is the same for each transition of a molecule.  To estimate $\Omega$ for each molecule, we use the deprojected radial profile of the brightest line (Figure \ref{fig_radprof}) to identify the maximum angular extent of emission.  $\Omega$ is taken to be the solid angle subtended by a circle with this radius.  In turn, the total column density $N_T$ and rotational temperature $T_{rot}$ can be determined from the upper level populations given by the Boltzmann distribution:

\begin{equation}
\label{eq_rd}
\frac{N_u}{g_u} = \frac{N_T}{Q(T_{rot})}e^{-E_u/T_{rot}}.
\end{equation}
\\
Here, $g_u$ is the upper state degeneracy, $Q$ is the molecular partition function, and $E_u$ is the energy of the upper state (K).  To calculate the partition functions for CH$_3$CN and HC$_3$N, we use the symmetric top and linear polyatomic approximations respectively \citep{Gordy1984,Mangum2015}:
\begin{eqnarray}
Q(T, \mathrm{CH_3CN}) = 1.78\times10^6\Big{(}\frac{T^3}{AB^2}\Big{)}^{1/2} \\
Q(T, \mathrm{HC_3N}) = \frac{kT}{hB_0}e^{hB_0/3kT}
\end{eqnarray}
where $A$, $B$, and $B_0$ are the rotational constants for CH$_3$CN and HC$_3$N.  

The best-fit CH$_3$CN and HC$_3$N column densities and rotational temperatures are listed in Table 3.  The rotational temperature of CH$_3$CN in V4046 Sgr is 29 $\pm$ 2K and in MWC 480 is 73 $\pm$ 23K.  By comparison, the rotational temperature of CH$_3$CN was recently measured in the disk around the solar analog TW Hya to be 29K; follow-up chemical modeling predicted abundant gas-phase CH$_3$CN between temperatures of $\sim$25--50K corresponding to the warm molecular layer of the disk \citep{Loomis2017}.  The measured rotational temperature in V4046 Sgr is consistent with these TW Hya results, while CH$_3$CN in MWC 480 is warmer.  MWC 480 is a Herbig Ae star and hosts a stronger radiation field than the T Tauri stars TW Hya and V4046 Sgr, which may explain the warmer emission.  However, we emphasize that given the line blending of CH$_3$CN (see Section \ref{sec_datareduce}) the rotational temperature is rather poorly constrained.  Moreover, at densities below $\sim$10$^7$ cm$^{-3}$, the CH$_3$CN and HC$_3$N transitions targeted in this survey will be sub-thermally excited.  Modeling in \citet{Loomis2017} suggests that CH$_3$CN emission can arise from regions with densities down to 10$^6$ cm$^{-3}$; therefore, the rotational temperatures listed in Table 3 may underestimate the kinetic temperature of the emission region.

\begin{deluxetable}{llll} 
	\tabletypesize{\footnotesize}
	\tablecaption{Best-fit column densities and rotational temperatures}
	\tablecolumns{4} 
	\tablewidth{\textwidth} 
	\tablehead{
		\colhead{Source}									&
		\colhead{Molecule}									&
		\colhead{N$_T$ (10$^{12}$ cm$^{-2}$)}					&
		\colhead{T$_{rot}$ (K)}								}
\startdata
MWC 480 & CH$_3$CN & 1.8 $\pm$ 0.4 & 73 $\pm$ 23 \\ 
V4046 Sgr & CH$_3$CN & 6.2 $\pm$ 1.8 & 29 $\pm$ 2 \\ 
MWC 480 & HC$_3$N & 5.8 $\pm$ 2.8 & 49 $\pm$ 6 \\ 
LkCa 15 & HC$_3$N & $>$0.8 & $<$103 \\ 
\enddata
\label{table_RD}
\end{deluxetable}

We note that in V4046 Sgr, the derived CH$_3$CN rotational temperature also corresponds to the kinetic gas temperature: for symmetric top molecules, the populations of different K levels with the same J value are a direct result of collisions \citep[e.g.][]{Loren1984}.  This is not the case for MWC 480 since the multiple lines detected consist of different J levels within the same K ladder.

HC$_3$N in MWC 480 has a measured rotational temperature of 49 $\pm$ 6K, consistent within the uncertainties with the warm CH$_3$CN temperature found in the same source.  Additional observations are needed to determine whether there is a real difference in the emission temperature (and therefore emission location) of CH$_3$CN and HC$_3$N within a disk.  In LkCa 15 only one HC$_3$N line was firmly detected, resulting in a rotational temperature upper limit of 103K.  

\subsection{Disk-averaged abundance ratios}
\label{sec_ratios}
Disk-averaged abundance ratios of HC$_3$N and CH$_3$CN with respect to HCN are calculated using the integrated fluxes listed in Table 2 and adopted rotational temperatures.  H$^{13}$CN integrated fluxes are taken from \citet{Guzman2017}; we assume a standard $^{12}$C/$^{13}$C ratio of 70 (with an uncertainty of 15\%) to convert to HCN column densities.  HC$_3$N/HCN and CH$_3$CN/HCN abundance ratios are calculated for rotational temperatures of 30, 50, and 70K corresponding to the range of observed rotational temperatures.  In this treatment we assume that HC$_3$N and CH$_3$CN are co-spatial with HCN.  If all molecules are emitting from the molecular layer between the midplane and the disk atmosphere this is a reasonable approximation vertically, however differences in the radial extent of emission for each molecule may introduce some error into the abundance ratios.  IM Lup is excluded from this analysis since only upper limits are available for all species.

The derived abundance ratios are listed in Table 4.  For all rotational temperatures, HC$_3$N is more abundant than CH$_3$CN.  The CH$_3$CN abundance is on the order of a few percent with respect to HCN for all choices of rotational temperature.  For a given rotational temperature, the CH$_3$CN/HCN ratios are consistent within the uncertainties with a single value across the entire disk sample.  The derived HC$_3$N/HCN abundance ratios are a few percent for a 70K rotational temperature but significantly higher ($\sim$50\% for AS 209, LkCa 15, and V4046 Sgr, and over 100\% for MWC 480 and HD 163296) for a 30K rotational temperature.  In MWC 480, the only source with a well-constrained HC$_3$N rotational diagram, the derived temperature is close to 50K; therefore, for at least the Herbig Ae stars, the 50K abundances ($\sim$20\%) are likely the most reliable.  

\begin{deluxetable*}{lllllll} 
	\tabletypesize{\footnotesize}
	\tablecaption{Disk-averaged CH$_3$CN/HCN and HC$_3$N/HCN abundance ratios}
	\tablecolumns{7} 
	\tablewidth{\textwidth} 
	\tablehead{
		\colhead{Source}									&
		\multicolumn{3}{c}{CH$_3$CN/HCN (\%)}					&
		\multicolumn{3}{c}{HC$_3$N/HCN (\%)}					\\
		\colhead{}											&
		\colhead{30K}										&
		\colhead{50K}										&
		\colhead{70K}										&
		\colhead{30K}										&
		\colhead{50K}        									&
		\colhead{70K}										}
\startdata
AS 209 & $<$ 4.7 &$<$ 2.5 &$<$ 2.0 &69.7 $\pm$ 17.7 &10.8 $\pm$ 2.8 &4.9 $\pm$ 1.2 \\ 
LkCa 15 & $<$ 4.3 &$<$ 2.3 &$<$ 1.8 &43.4 $\pm$ 13.2 &6.7 $\pm$ 2.0 &3.0 $\pm$ 0.9 \\ 
V4046 Sgr & \textbf{5.1 $\pm$ 1.8} &2.7 $\pm$ 0.9 &2.2 $\pm$ 0.8 &37.9 $\pm$ 6.2 &5.9 $\pm$ 1.0 &2.7 $\pm$ 0.4 \\ 
HD 163296 & 5.6 $\pm$ 1.6 &2.9 $\pm$ 0.9 &2.4 $\pm$ 0.7 &134.1 $\pm$ 28.0 &20.9 $\pm$ 4.4 &9.4 $\pm$ 2.0 \\ 
MWC 480 & 5.9 $\pm$ 1.9 &3.1 $\pm$ 1.0 &\textbf{2.5 $\pm$ 0.8} &121.9 $\pm$ 24.0 &\textbf{19.0 $\pm$ 3.7} &8.5 $\pm$ 1.7 \\ 
\enddata
\tablenotetext{}{For disks with measured rotational temperatures, the closest corresponding abundance is marked in bold.}
\label{table_ratios}
\end{deluxetable*}

\section{Abundance profile modeling}
\label{sec_model}
The strong emission of both CH$_3$CN and HC$_3$N in MWC 480 and V046 Sgr enables a more detailed modeling of the radial abundance profiles within each disk.

\subsection{V4046 Sgr model}
\label{sec_vmodel}
The physical model for V4046 Sgr is adapted from the parametric model of the V4046 Sgr disk described in \citet{Rosenfeld2013} and is further developed in \citet{Guzman2017}.  For fitting CH$_3$CN and HC$_3$N emission, we use the same physical disk model as described in \citet{Guzman2017}. 

Molecular abundance profiles are assumed to follow a power law, following e.g. \citet{Qi2008,Qi2013}: 
\begin{equation}
X (r)= X_{100} \Big{(}\frac{r}{R_{100}}\Big{)}^\alpha,
\label{eq_abud}
\end{equation}
where $X$ is the abundance with respect to the total hydrogen density, $X_{100}$ is the abundance at the characteristic radius of $R_{100}$ = 100 AU, and $\alpha$ is the power law index.  An outer radius cutoff $R_{out}$ of 100AU was adopted based on the extent of emission in the deprojected radial profile (Figure \ref{fig_radprof}).  In the disk atmosphere, photo-dissociation is assumed to destroy most molecules, and the molecular abundances are attenuated by a factor of 10$^{8}$ above $z/r$ = 0.5.  To account for depletion in the disk midplane, molecule abundances are attenuated by a factor of 10$^{3}$ at temperatures below 25K.  This temperature does not correspond to a purely thermal freeze-out boundary for nitriles, but was chosen empirically based on the boundary where gas-phase CH$_3$CN disappears in the chemical model presented in \citet{Loomis2017}, and is also consistent with the $\sim$30K rotational temperature found for CH$_3$CN in this disk.  This depletion boundary is similar to the expected CO freeze-out temperature, and likely arises due to either a coincidence with the photodesorption boundary of the nitrile molecules, or an increase in gas-phase nitrile chemistry driven by CO sublimation.

CH$_3$CN and HC$_3$N were fit independently for the free parameters $X_{100}$ and $\alpha$.  While we use the same physical disk model as in \citet{Guzman2017}, the boundary conditions for the molecular abundance profiles are slightly different here; since we ultimately wish to normalize CH$_3$CN and HC$_3$N with respect to H$^{13}$CN, we also re-fit the H$^{13}$CN observations to ensure consistency.

\begin{figure*}
	\includegraphics[width=\linewidth]{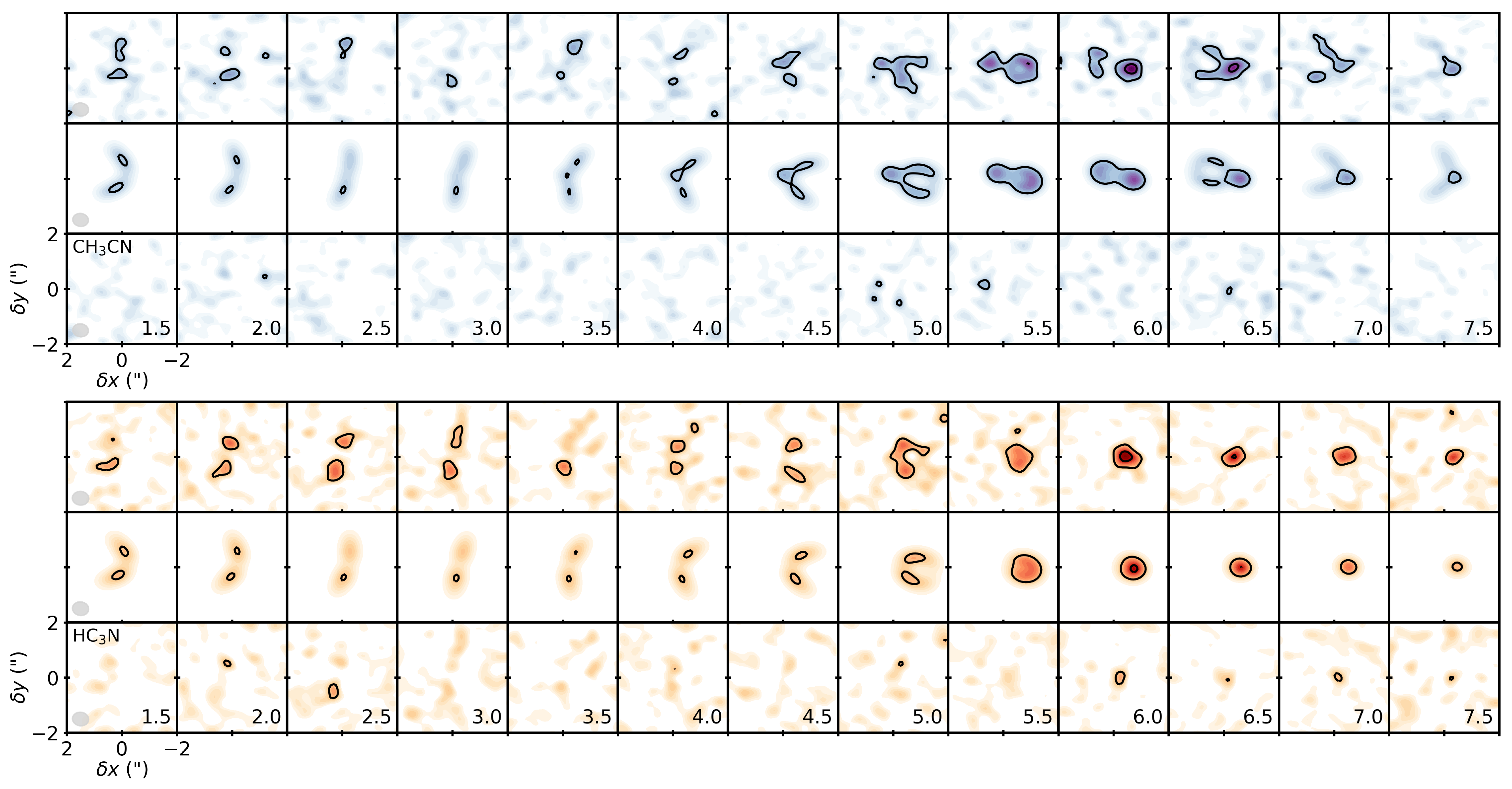}
	\caption{Channel maps of observations (top rows), best-fit models (middle rows), and residuals (bottom rows) for CH$_3$CN 14$_0$-13$_0$ + 14$_1$-13$_1$ and HC$_3$N 27-26 in V4046 Sgr.  Contour levels correspond to 3, 5, and 7$\times$ rms.  The velocity (km/s) is indicated in the bottom row for each molecule, and the synthesized beam is shown in the first panel of each row.}
\label{fig_modelvobs}
\end{figure*}

The fitting was performed by generating synthetic observations of the brightest line emission, holding the gas density and temperature profiles constant and adopting the disk inclination, position angle, stellar mass, and systemic velocity listed in Table 1.  The synthetic images had a spectral resolution of 0.5 km s$^{-1}$ and spanned -2.0 -- 14.5 km/s for CH$_3$CN (including both the 14$_0$--13$_0$ and 14$_1$--13$_1$ transitions) and -3.0 -- 9.0 km/s for HC$_3$N.  The radiative transfer code {\fontfamily{qcr}\selectfont RADMC-3D} \citep{Dullemond2012} was used to calculate level populations for each synthetic image assuming LTE conditions.  The {\fontfamily{qcr}\selectfont vis\textunderscore sample} Python package\footnote{https://pypi.python.org/pypi/vis\_sample} \citep{Loomis2017a} was then used to sample the synthetic image at the $u-v$ points of the observations.  The likelihood function was calculated from the weighted difference between observations and the model in the $u-v$ plane.  The affine-invariant MCMC package {\fontfamily{qcr}\selectfont emcee} \citep{Foreman-Mackey2013} was used to sample posterior distributions of both parameters $X_{100}$ and $\alpha$.  A flat prior was used for each parameter when generating new samples: $10^{-20}  < X_{100} <$ 10$^{-8}$ and -3 $< \alpha <$ 2.  The resulting best-fit values are listed in Table 5.  Figure \ref{fig_modelvobs} shows channel maps of the observations along with the best-fit model and residuals for CH$_3$CN and HC$_3$N (H$^{13}$CN and CH$_3$CN 14$_2$--13$_2$ can be found in Appendix \ref{app_models}); the observations are well reproduced by the model.

To ensure that the choice of depletion boundary does not significantly impact our results, the models were also run with an adopted depletion boundary of 19K and 30K.  For CH$_3$CN the best-fit $X_{100}$ at 19K and 30K are within 25\% of the 25K value, and the best-fit $\alpha$ are within 15\%.  For HC$_3$N there was less than 3\% change for both $X_{100}$ and $\alpha$.  The results are therefore not highly sensitive to the choice of depletion boundary.  To further confirm the derived abundances we use the best-fit $X_{100}$ and $\alpha$ values from the CH$_3$CN 14$_0$--13$_0$ line to create a synthetic image of the 14$_2$--13$_2$ transition.  The higher-frequency transition is well reproduced by the lower-frequency best-fit values (shown in Appendix \ref{app_models}), indicating that the adopted model is appropriate.

Since V4046 Sgr is a transition disk with a large inner gap, we also test whether a model with a large cavity produces a better fit to the observations.  The fiducial model uses a cavity radius of 3 AU based on the CO line emission and SED of V4046 Sgr \citep{Rosenfeld2013}.  An adopted 29 AU radius, corresponding to the mm dust radius hole, produces a worse fit to the data for both molecules.

\begin{deluxetable*}{lrrrr} 
	\tabletypesize{\footnotesize}
	\tablecaption{Abundance modeling best-fit parameters}
	\tablecolumns{5} 
	\tablewidth{\textwidth} 
	\tablehead{
		\colhead{}											&
		\multicolumn{2}{c}{V4046 Sgr}					 		&
		\multicolumn{2}{c}{MWC 480}							\\
		\colhead{}											&
		\colhead{$X_{100}$}									&
		\colhead{$\alpha$}									&
		\colhead{$X_{100}$}									&
		\colhead{$\alpha$}									}
\startdata
CH$_3$CN  & 8.1 $\pm$ 0.4 $\times$ 10$^{-12}$ & 0.6 $\pm$ 0.1 & 
			7.9 $^{+2.7}_{-2.1} \times$ 10$^{-13}$ & -0.1 $\pm$ 0.3 \\
HC$_3$N     &  2.8 $\pm$ 0.2 $\times$ 10$^{-11}$ & 1.3 $\pm$ 0.1 & 
			1.2 $\pm$ 0.1 $\times$ 10$^{-11}$ & 0.8 $\pm$ 0.1 \\
H$^{13}$CN & 1.2 $\pm$ 0.1 $\times$ 10$^{-12}$ & -0.5 $\pm$ 0.1 & 
			1.2 $\pm$ 0.2 $\times$ 10$^{-13}$ & -1.1 $\pm$ 0.1 \\
\enddata
\label{model_bestfits}
\end{deluxetable*}

\subsection{MWC 480 model}
\label{sec_mmodel}
In \citet{Oberg2015}, the CH$_3$CN 14$_0$--13$_0$ and HC$_3$N 27--26 profiles in MWC 480 were fit by obtaining the minimum $\chi^2$ value from a grid of calculated abundance models.  Here, we adopt the same parametric physical model for the disk density and temperature described in that paper but use the MCMC fitting procedure described in the previous section to constrain $X_{100}$ and $\alpha$.  This allows us to better explore the parameter space and therefore to obtain more robust constraints on the fit parameters and their uncertainties.  Also in contrast to \citet{Oberg2015}, we use {\fontfamily{qcr}\selectfont RADMC-3D} instead of the non-LTE {\fontfamily{qcr}\selectfont LIME} code for radiative transfer calculations to ensure consistency with the V4046 Sgr results.  

To retrieve molecular abundances, we use the same power-law prescription (Equation 5) as in Section \ref{sec_vmodel}.  Again the abundances are attenuated above a $z/r$ of 0.5 to account for photo-destruction.  We adopt a depletion boundary of $z/r <$ 0.05 which roughly corresponds to the cutoff in the V4046 Sgr model.  As in the V4046 Sgr model, this does not correspond to the nitrile freeze-out boundary but likely corresponds to where photodesorption of nitriles and/or CO-driven gas phase chemistry become efficient.  A higher cutoff of $z/r <$ 0.2 was also tested, corresponding to the lower boundary of CH$_3$CN emission in the model of \citet{Loomis2017}.  An outer radius $R_{out}$ of 180 AU was chosen, corresponding to the extent of emission in the MWC 480 radial profiles (Figure \ref{fig_radprof}).  The MCMC fitting procedure is otherwise the same as described above.  CH$_3$CN, HC$_3$N, and H$^{13}$CN were each fit with spectral resolutions of 0.5km/s and velocity ranges of 1--9.5km/s, 0--14.5km/s, and 0--14.5km/s, respectively.  The resulting model channel maps and residuals are shown in Appendix \ref{app_models}.  The best-fit values of X$_{100}$ and $\alpha$ are listed in Table 5.  For both V4046 Sgr and MWC 480, CH$_3$CN and HC$_3$N appear to have increasing or flat abundance profiles, while H$^{13}$CN shows a decreasing abundance profile.  

Like for V4046 Sgr we can use the constraints from additional lines to confirm the validity of the best-fit model.  For both depletion boundaries of $z/r < 0.05$ and $z/r < 0.2$ we created synthetic images of the upper-level CH$_3$CN and HC$_3$N lines based on the best-fit $X_{100}$ and $\alpha$ values from the lower-level lines.  Rotational diagram calculations were performed for CH$_3$CN and HC$_3$N using the modeled fluxes.  We obtain very similar rotational temperatures for $z/r < 0.05$ and $z/r < 0.2$ models: 20K and 23K for CH$_3$CN, and 44K and 45K for HC$_3$N, respectively.  The HC$_3$N model temperatures are very close to the observed values, while the CH$_3$CN temperatures are low.  Because the modeled CH$_3$CN rotational temperature is not substantially improved by increasing the $z/r$ cutoff, a more complex parametric model is likely required to fully describe the disk physical and/or abundance structure.  For instance, modeling of H$_2$CO emission in TW Hya required both a hot inner component and a cool extended component to match observations \citep{Oberg2017}; the possible presence of a warm inner component in MWC 480 would not be captured by the single power-law profile used in our models, resulting in a potential under-prediction of the observed rotational temperature.  We emphasize, however, that the observed rotational temperature of CH$_3$CN in MWC 480 is not well constrained due to a small lever arm in upper energy levels as well as line blending uncertainties, and may in reality be closer to the modeled rotational temperature.  Of the two depletion boundaries, the $z/r < 0.05$ cutoff produced better fits to the higher-J lines as determined by the reduced $\chi^2$, and therefore all future discussion pertains to the $z/r <$ 0.05 model results. Channel maps for these models are shown in Appendix \ref{app_models}.

\begin{figure}
	\includegraphics[width=\linewidth]{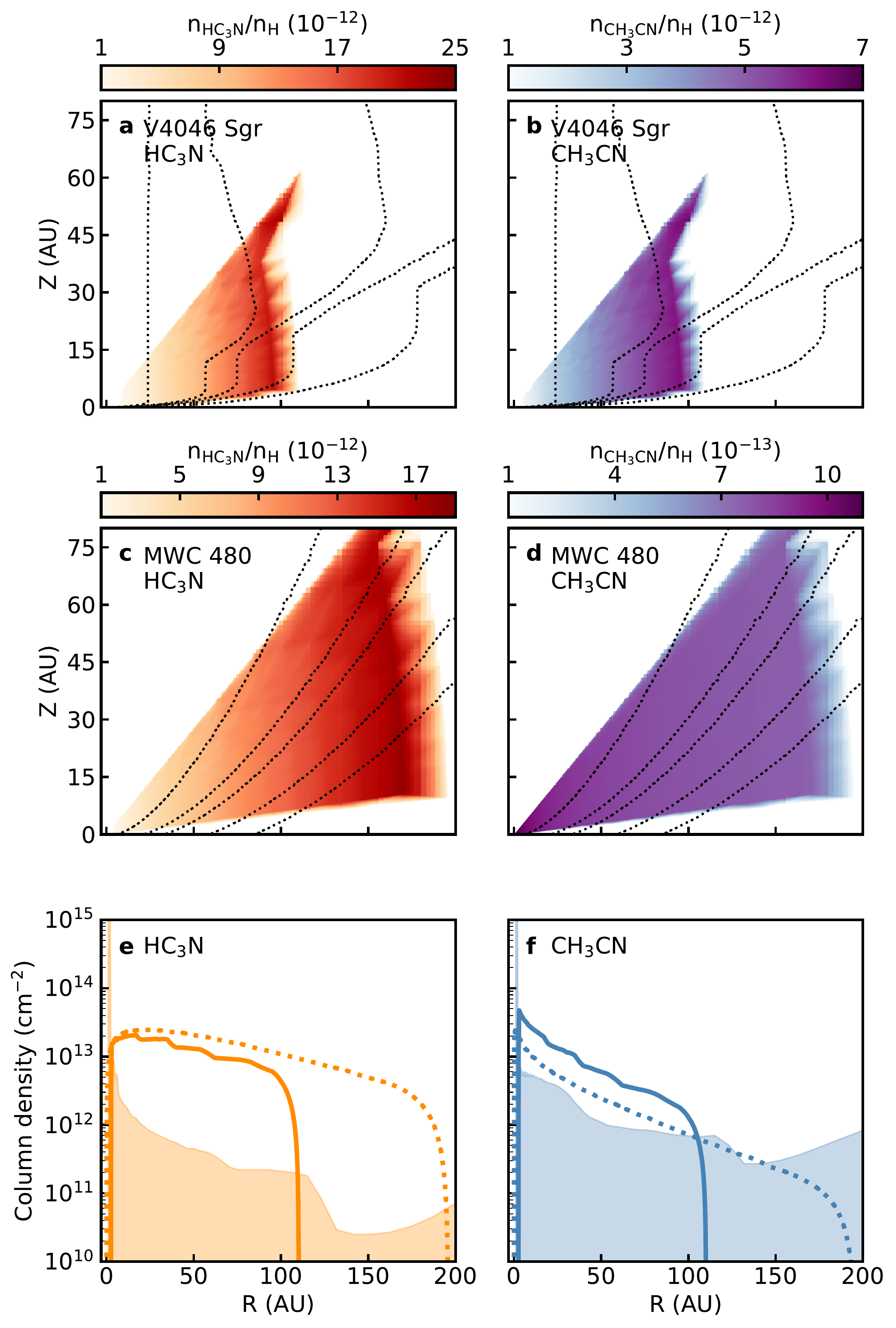}
	\caption{Parametric abundance modeling results.  (a--d) Best-fit abundance profiles of HC$_3$N and CH$_3$CN with respect to total H density in V4046 Sgr and MWC 480.  Dotted lines indicate 20, 30, 40, 50, and 100K temperature contours.  (e,f) Derived column densities as a function of radius for HC$_3$N and CH$_3$CN; solid and dotted lines represent V4046 Sgr and MWC 480 respectively.  Shaded regions show column densities resulting from the disk chemistry model in \citet{Walsh2014}.}
\label{fig_diskmodel}
\end{figure}

\subsection{CH$_3$CN and HC$_3$N column densities and abundances}
\label{sec_abundCD}
The best-fit CH$_3$CN and HC$_3$N abundance profiles derived for V4046 Sgr and MWC 480 are shown in Figure \ref{fig_diskmodel}, along with the derived radial column density profiles.  For comparison, we also show the column densities of HC$_3$N and CH$_3$CN predicted by a disk chemistry model for a generic T Tauri star and disk \citep{Walsh2014}.  CH$_3$CN column densities from the disk chemistry model are within an order of magnitude to those derived in this work for V4046 Sgr and MWC 480, while HC$_3$N column densities are under-estimated by over an order of magnitude in the model.  However, neither V4046 Sgr nor MWC 480 is well-described by the disk physical structure adopted by \citet{Walsh2014}, and further tuning of models is required to make conclusive comparisons.  Comparing the relative shapes of the radial column density profiles, the extremely centrally peaked profile in the model is not reproduced in the profiles derived in this work.  However, we note that due to the high upper-state energies of the lines fitted (92K for CH$_3$CN and 165K for HC$_3$N), our observations are not sensitive to cool material and therefore may not reflect the true spatial distributions of the molecules. 

\begin{figure}
	\includegraphics[width=\linewidth]{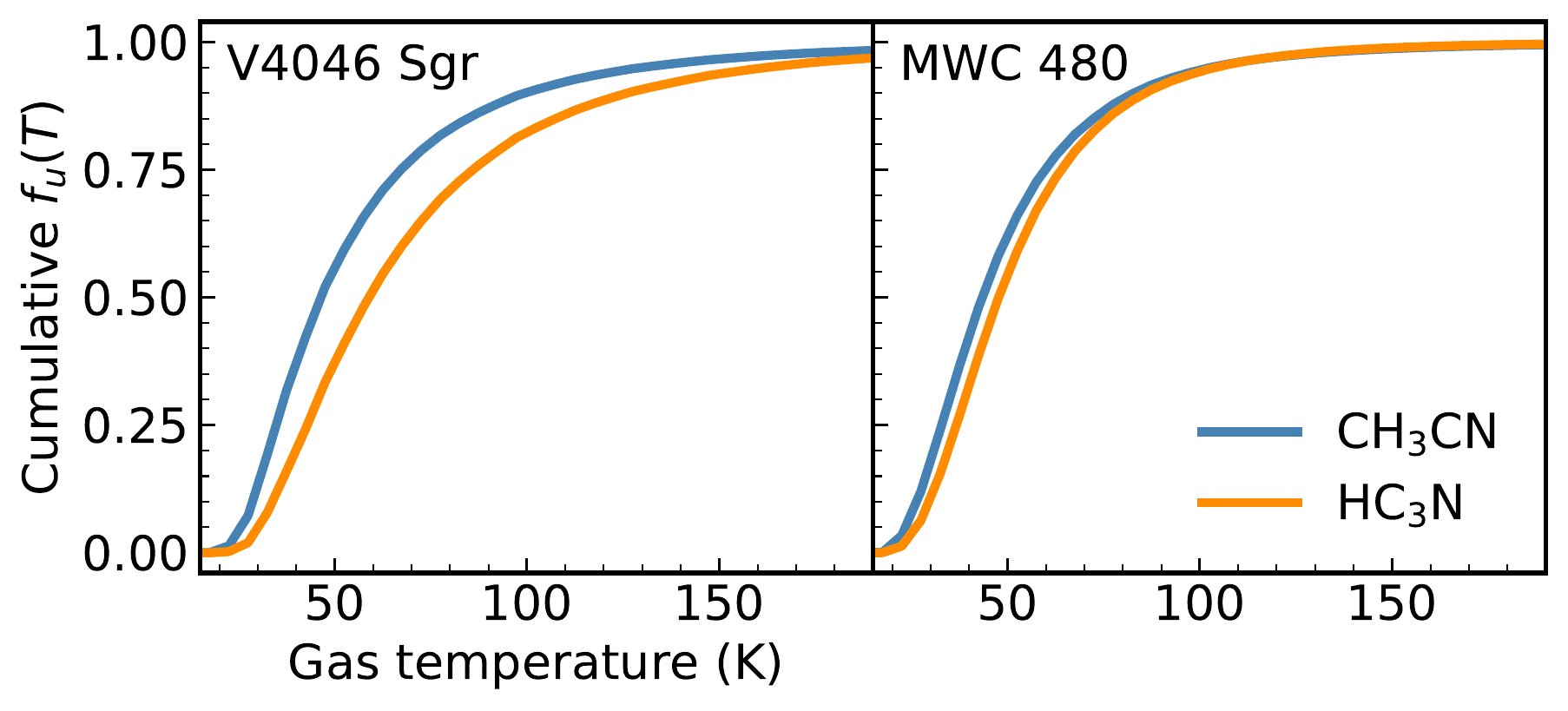}
	\caption{Model-derived cumulative fraction of molecules in the upper energy state of the observed transition, as a function of increasing gas temperature.}
\label{fig_tempbins}
\end{figure}

To illustrate this, we use the modeled best-fit abundance profiles to determine the fraction of emitting molecules (i.e., molecules in the upper energy state of the observed transition) in temperature bins from 10 to 200K, following \citet{Bergin2013}.   The number density of a species in the upper energy state $n_u$ is related to the total number density $n_T$ by the Boltzmann distribution (Equation \ref{eq_rd}).  $n_u$ is integrated over the disk to find the number of upper-state molecules in target temperature bins.  The fraction of upper-state molecules in each temperature bin relative to the total number of upper-state molecules, $f_u(T)$, is shown in Figure \ref{fig_tempbins} as a cumulative distribution function.

For both CH$_3$CN and HC$_3$N in V4046 Sgr and MWC 480, roughly half of the emitting molecules are in gas warmer than 50K, with virtually no contribution from 30K gas.  This indicates that our observations are mostly probing warm emission.  Since most gas in disks exists at $<$50K temperatures, there may be a substantial amount of material that our observations are not sensitive to.  As further discussed in Section \ref{sec_discussion}, follow-up observations of lower-J transitions of HC$_3$N and CH$_3$CN will be helpful in addressing this issue.

Because the retrieved abundance profiles may depend on the physical disk structure assumed in the model, we also present the CH$_3$CN and HC$_3$N abundance profiles normalized with respect to HCN.  This should be less sensitive to the details of the physical model structure assuming that all three molecules are emitting co-spatially; this is already implicit in our model since we used the same  freeze-out and photo-dissociation boundary conditions to retrieve molecular abundances within a given source.  H$^{13}$CN is converted to HCN using the standard isotopic ratio of 70 with an uncertainty of 15\%.

\begin{figure}
	\includegraphics[width=\linewidth]{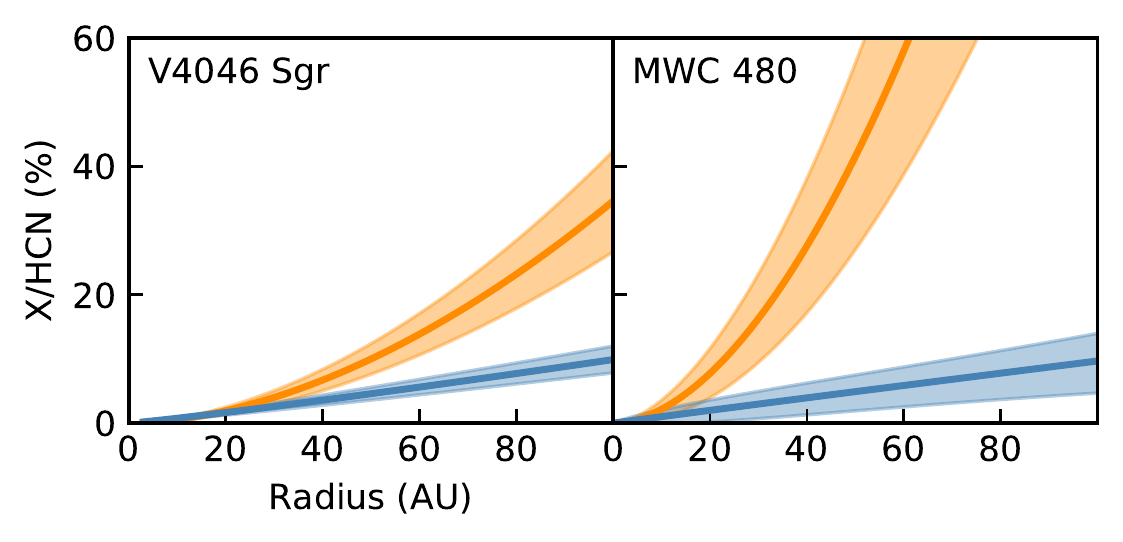}
	\caption{Best-fit abundance profiles with respect to HCN for CH$_3$CN (blue) and HC$_3$N (orange) in V4046 Sgr and MWC 480.  Dark lines show the best-fit model for each molecule, with the lighter ribbons representing uncertainties. 
	}
\label{fig_abundprof}
\end{figure}

The resulting CH$_3$CN and HC$_3$N abundance profiles with respect to HCN in V4046 Sgr and MWC 480 are shown in Figure \ref{fig_abundprof}.   The derived gas-phase abundances with respect to HCN in the inner 100AU of the disks are on the order of a few percent for CH$_3$CN and a few tens of percent for HC$_3$N.  These model-derived inner disk abundances are consistent with the range of disk-averaged abundances calculated assuming 30--70K rotational temperatures (Table 4).  

\section{Discussion}
\label{sec_discussion}
In a sample of 6 protoplanetary disks, we have detected the complex nitrile molecules HC$_3$N and CH$_3$CN in five and three disks respectively.  These molecules therefore appear common in other nascent planeatry systems.  The disks in our sample host a range of physical conditions, which can be used to evaluate the nitrile chemistry in disks.  We begin by surveying the possible origins of complex nitriles in disks, followed by an evaluation of our source sample.

\subsection{Nitrile formation in disks}
\subsubsection{Chemical pathways}
\label{sec_chem}
HC$_3$N is proposed to form efficiently in the gas phase, via either CN + C$_2$H$_2$ or HCN + C$_2$H \citep{Fukuzawa1997}, and has no known efficient grain surface formation pathways.  In contrast, CH$_3$CN can form via both gas-phase and grain-surface processes.  In current astrochemistry codes, the dominant gas phase CH$_3$CN formation channel is CH$_3^+$ + HCN followed by dissociative recombination.  On grain surfaces, CH$_3$ + CN or hydrogenation of C$_2$N are proposed to be efficient at forming CH$_3$CN \citep{Huntress1979,Walsh2014}.  Current evidence suggests that grain-surface chemistry is a primary contributor of CH$_3$CN in disk environments: in \citet{Oberg2015}, a gas-phase only chemical model failed to reproduce observed abundances of CH$_3$CN/HCN towards MWC 480, implying a significant contribution from grain-surface chemistry to the observed gas-phase abundances.  Likewise, \citet{Loomis2017} test a gas-phase only and gas-grain model and find that grain chemistry is required to reproduce observed CH$_3$CN abundances in the disk around TW Hya.   We note that all of the proposed formation mechanisms of CH$_3$CN have estimated rate constants and have not been experimentally validated, and there may be important contributions from as yet unexplored chemistry that leads to CH$_3$CN formation.  

\subsubsection{Nitrile abundance correlations}

\begin{figure}
	\includegraphics[width=\linewidth]{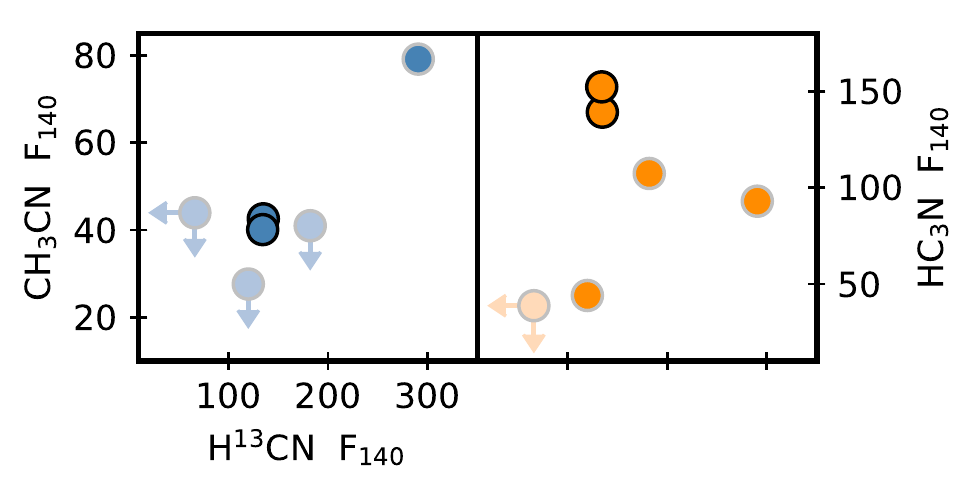}
	\caption{Integrated fluxes (normalized to a distance of 140 pc) for CH$_3$CN 14$_0$--13$_0$ and HC$_3$N 27--26 plotted against H$^{13}$CN 3--2.  Silver and black marker borders indicate T Tauri and Herbig Ae disks respectively.  Upper limits are shown as faint colors, and units for all axes are mJy km s$^{-1}$.}
\label{fig_vshcn}
\end{figure}

To explore the relationship among different nitrile-bearing species in our disk sample, Figure \ref{fig_vshcn} shows the distance-normalized fluxes of the CH$_3$CN 14$_0$--13$_0$ and HC$_3$N 27--26 transitions each plotted against the H$^{13}$CN 3--2 transition.

The HC$_3$N emission strength has no clear relation to H$^{13}$CN.  However, interpreting this lack of correlation is complicated by the high upper energy of the HC$_3$N 27--26 line: with an excitation temperature of 165K, this transition is not sensitive to cool HC$_3$N molecules which may be abundant in some disks.  Indeed, the enhanced HC$_3$N emission around the Herbig Ae stars compared with the T Tauri stars is consistent with a thermal effect (Figure \ref{fig_vshcn}), as there will be more hot molecular material around more luminous stars.  Observations of lower-J HC$_3$N transitions are needed to determine whether the HC$_3$N chemistry is related to the other nitrile chemistry in disks.

From the available data, CH$_3$CN emission appears to correlate with H$^{13}$CN, although more detections are required to establish if this relationship is real.  We note that CH$_3$CN exhibits no such correlation with C$^{18}$O emission strength; the tentative correlation with H$^{13}$CN is therefore not simply a trend with the amount of gas in the disk.  If additional data points confirm this correlation, it may be evidence for an active gas-phase contribution to CH$_3$CN formation, as this is currently the only chemical pathway with a direct link between HCN and CH$_3$CN.  Other potential gas-phase and grain-surface channels to CH$_3$CN formation which could explain a correlation with HCN but are not currently included in models should be also explored.  

\subsubsection{Nitrile spatial correlations}
Correlations in the spatial extent of molecules within a disk can also be used to constrain their formation chemistry.  Across the disk sample, the spatial distributions of CH$_3$CN and HC$_3$N (Figure \ref{fig_mom0spec_main}) as well as H$^{13}$CN  \citep[see] [for H$^{13}$CN maps]{Guzman2017} are all compact, typically well within the bounds of the dust continuum.  This spatial similarity of nitrile emission within each disk is consistent with a chemical scheme in which CH$_3$CN and HC$_3$N depend on abundant HCN (or its photo-product CN) to form.  We note that this is a stronger constraint for CH$_3$CN and H$^{13}$CN than for HC$_3$N because, as discussed above, the emission from high-J transitions of HC$_3$N may not reflect the true distribution of molecules within the disk, and lower-energy transitions are required to confirm a compact distribution.  

Determining whether the spatial distributions of HC$_3$N and its proposed precursor C$_2$H are related will provide important constraints on the HC$_3$N formation chemistry.  Spatially resolved observations of C$_2$H towards TW Hya show a ringed structure peaking near the edge of the sub-millimeter continuum; DM Tau similarly demonstrates an outer ring near the dust edge and an inner ring co-spatial with the continuum \citep{Kastner2015, Bergin2016}.  A ringed morphology may be a feature of hydrocarbons more generally, and indeed is reproduced for the slightly larger hydrocarbon C$_3$H$_2$ \citep{Bergin2016, Qi2013}.  The apparent anti-correlation of HC$_3$N and C$_2$H suggests that the proposed HC$_3$N formation pathway of C$_2$H + HCN may not be efficient.  If the ringed morphology is common to all hydrocarbons, this is also problematic for the C$_2$H$_2$ + CN pathway, although the relationship between C$_2$H and C$_2$H$_2$ distributions is unconstrained.  Yet again we note that due to the high-energy HC$_3$N transitions and comparatively low angular resolution of our observations we cannot exclude the possibility of HC$_3$N rings.  Higher-resolution observations of lower-J HC$_3$N transitions combined with C$_2$H observations in the same sources will be important for constraining whether the current HC$_3$N formation paradigm is viable.

\subsection{Physical drivers of nitrile chemistry in disks}
The physical conditions of a protoplanetary disk set what chemistry can occur; while myriad properties have been proposed as chemical drivers in disks, we focus here on those which are to some degree testable based on the properties of our sample, namely the radiation field, disk structure, and evolutionary stage.  

\subsubsection{Radiation field}
\begin{figure*}
	\includegraphics[width=0.8\linewidth]{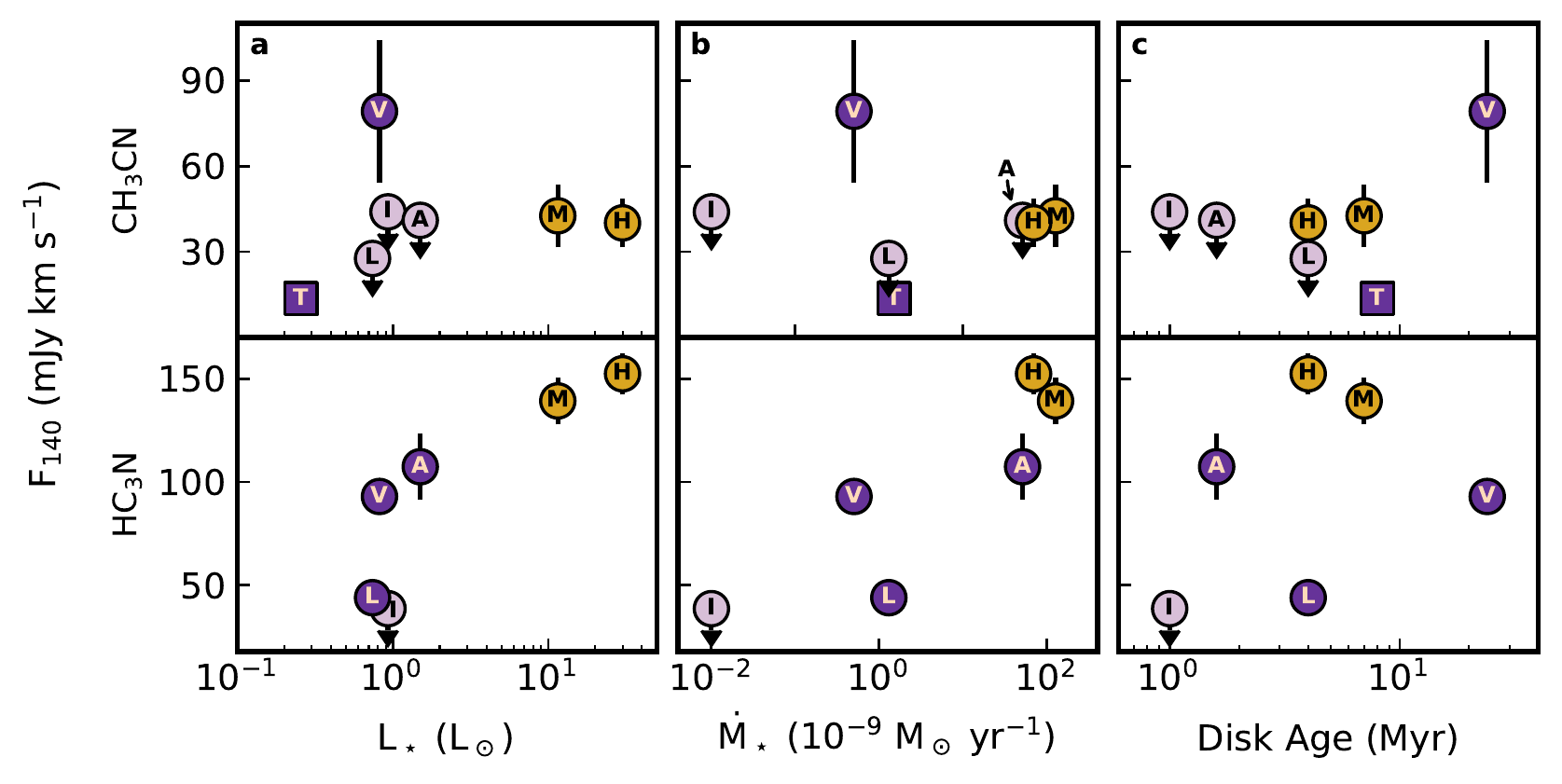}
	\caption{Integrated fluxes (normalized to a distance of 140 pc) for CH$_3$CN 14$_0$--13$_0$ and HC$_3$N 27--26, plotted against the quiescent luminosity, mass accretion rate, and disk age of the host star.  Disks are identified by the first letter of their name.  Purple and gold markers represent T Tauri and Herbig Ae stars, respectively.  Circle markers are disks observed in this survey; the square marker represents the predicted flux in TW Hya based on \citet{Loomis2017}.  3$\sigma$ upper limits are shown with faint colors. }
\label{fig_vsuv}
\end{figure*}

The quiescent luminosity and accretion luminosity of a host star both contribute to the overall radiation environment in a disk.  The FUV emission in T Tauri stars arises dominantly from accretion luminosity, while Herbig Ae stars should have significant FUV contributions from quiescent stellar photospheric emission in addition to accretion luminosity \citep[e.g.][]{Kurucz1993, Matsuyama2003}.  Figure \ref{fig_vsuv}a--b shows the distance-normalized disk-integrated fluxes of CH$_3$CN 14$_0$--13$_0$ and HC$_3$N 27--26 plotted against the quiescent luminosity and the mass accretion rate of each star.  For comparison, the CH$_3$CN 14$_0$--13$_0$ flux calculated for TW Hya based on the observations of \citet{Loomis2017} is also included, with a bolometric luminosity and mass accretion rate taken from \citet{vanBoekel2017} and \citet{Herczeg2008} respectively.  For CH$_3$CN we do not see any obvious trends with L$_\star$ or $\dot{M}_\star$.  This lack of correlation with the radiation field could indicate emission from the colder UV-shielded layers of the disk, but is also possibly due to the small number of CH$_3$CN detections.  HC$_3$N appears to correlate with both the stellar luminosity and the mass accretion rate, suggesting that the UV field may play an important role in driving its chemistry.  However, due to the high upper energy of the 27--26 transition, this could be due in part to the presence of hotter gas in high-UV environments rather than increased abundances of HC$_3$N; observations of lower-J HC$_3$N lines will be able to break this degeneracy.
 
\subsubsection{Disk age}
As disks evolve, processes such as viscous accretion, dust growth/settling, and radial drift reshape the physical structure of the disk \citep[reviewed in][]{Williams2011}.  Astrochemical modelers have recently begun to explore how a dynamically evolving disk impacts the chemistry, with a particular focus on the C/O ratio over time \citep{Piso2015, Eistrup2017}.  Modeling by \citet{Du2015} shows that the abundance of nitrile species can be greatly enhanced as a result of gas-phase carbon and oxygen depletion: as the system ages and more CO and H$_2$O are depleted from the gas-phase, the nitrile abundances should correspondingly increase.  Observationally, \citet{Kastner2014} observed enhanced CN abundances towards the evolved disks around TW Hya and V4046 Sgr.  

Figure \ref{fig_vsuv}c shows the CH$_3$CN 14$_0$--13$_0$ and HC$_3$N 27--26 integrated fluxes normalized to a distance of 140pc and plotted against the disk age.  Again, the CH$_3$CN 14$_0$--13$_0$ flux in TW Hya calculated from the observations of \citet{Loomis2017} is included for comparison.  V4046 Sgr, the oldest disk in the sample, shows anomalously high CH$_3$CN emission.  However, in all other disks CH$_3$CN detections and upper limits are fairly clustered, showing no obvious trend with age.  Likewise, HC$_3$N emission does not appear related to disk age.  With the existing data there is therefore insufficient evidence for an evolutionary trend in nitrile emission.  We note that the discrepancy in CH$_3$CN emission between V4046 Sgr and TW Hya is somewhat surprising given that the line intensities of other small molecules in the two disks are quite similar \citep{Kastner2014}.  

\subsubsection{Inner dust cavity}
The disk structure sets set how radiation is processed through the disk.  Transitional disks, characterized by inner gaps in mm dust emission, may host a distinct chemistry due to increased UV radiation in the inner disk \citep[e.g.][]{Cleeves2011}.  LkCa 15 and V4046 Sgr are both transition disks and yet exhibit very different nitrile chemistries: V4046 Sgr is strongly detected in both CH$_3$CN and HC$_3$N, while LkCa 15 is weakly detected in HC$_3$N and tentatively or not detected in CH$_3$CN.  There is therefore no strong global impact of an inner cavity on the disk's nitrile chemistry; observations towards other transition disks are needed to confirm this in a larger sample.  On smaller scales, we expect that the presence of an inner gap would result in warmer gas and a higher UV field within the cavity.  Suggestively, there is a slight peak in the radial profile of CH$_3$CN in LkCa 15 out to $\sim$50 AU scales (Figure \ref{fig_radprof}), consistent with the cavity radius derived by \citet{Pietu2006}; however, this emission is not significant at the 3$\sigma$ level and therefore no firm conclusions can be drawn.

\subsection{Nitriles in different circumstellar environments}
\label{sec_compare}
We now compare the disk-averaged abundance ratios for our sample with the abundances measured in similar objects at different evolutionary stages.  Low-mass protostars are the evolutionary precursors to the $<$2M$_\odot$ stars in this sample, while comets formed out of the midplane of the protosolar nebula and should preserve material from the time of planet formation.  We note the environments in these different types of objects span a wide range of temperatures, densities, radiation fields, and other physical conditions.

As discussed in Section \ref{sec_abundCD}, for V4046 Sgr and MWC 480 the model-derived CH$_3$CN/HCN and HC$_3$N/HCN abundances in the inner 100 AU are consistent with the range of disk-averaged abundances calculated for 30K--70K rotational temperatures.  In this section, we therefore use the range of disk-averaged abundances as representative of the inner 100AU of the disk in order to compare across the entire disk sample.

\begin{figure}
	\includegraphics[width=0.9\linewidth]{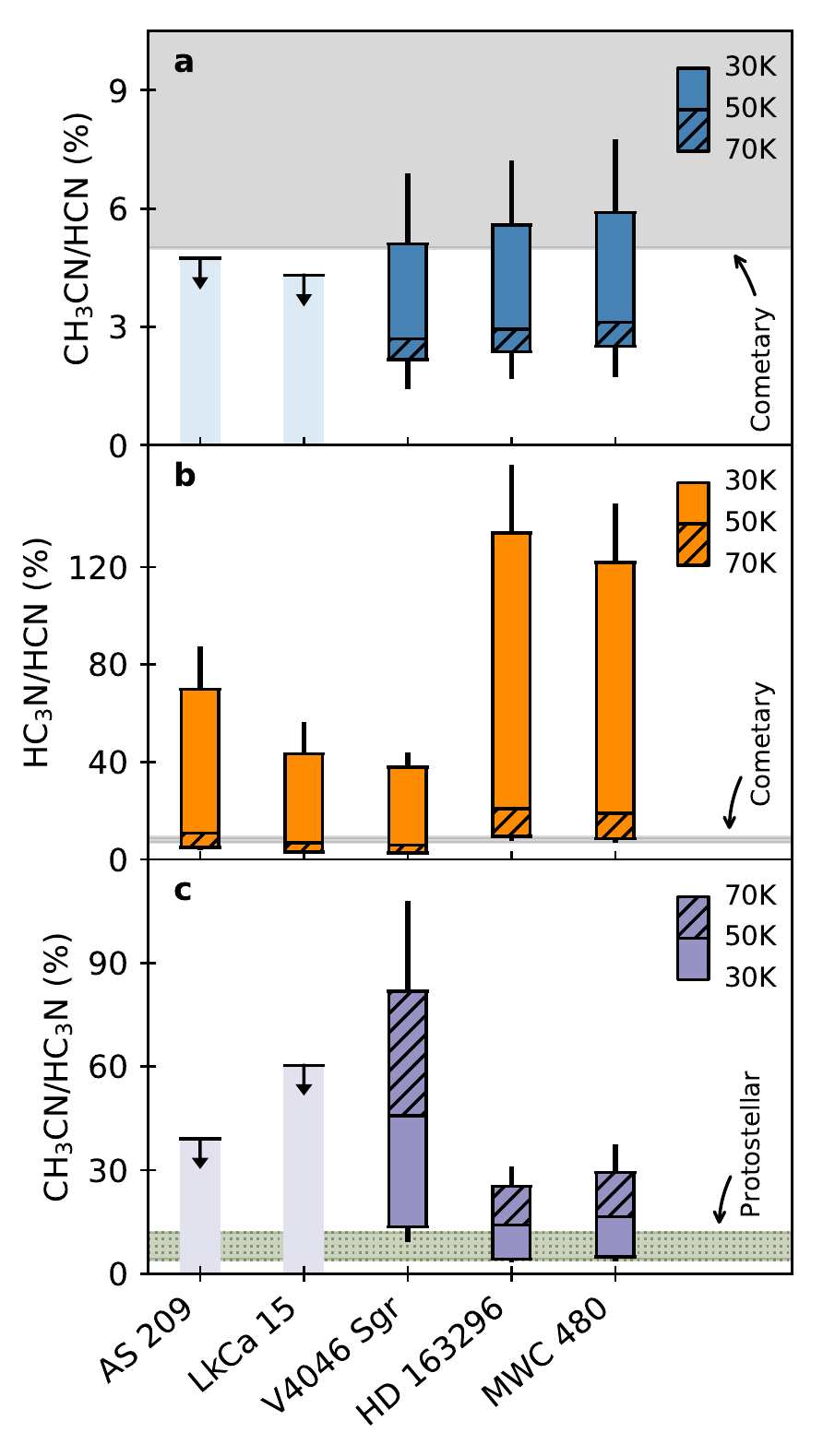}
	\caption{Disk-averaged ratios of (a) CH$_3$CN/HCN, (b) HC$_3$N/HCN, and (c) CH$_3$CN/HC$_3$N.  For each source, colored bars span the range of ratios calculated assuming 30--70K rotational temperatures.  The 30--50K and 50--70K ranges are represented as solid and dashed bars respectively, as indicated in the legend.  Error bars represent the uncertainty calculated at each bounding value.  In panels a--b the range of ratios measured in comets \citep{Mumma2011} is shown as a grey band; in panel c, the range of values measured in a sample of protostars \citep{Bergner2017} is shown as a dotted green band.}
\label{fig_ratios}
\end{figure}

Figure \ref{fig_ratios}a--b show the range of CH$_3$CN/HCN and HC$_3$N/HCN abundances measured in solar system comets \citep{Mumma2011} compared to the gas-phase abundances measured in our disk sample.  The disk abundances of CH$_3$CN are within a few percent of the values measured in solar system comets for sources with detections.  The upper limits for AS 209 and LkCa 15 are somewhat lower but still possibly within a few percent of cometary.  

For HC$_3$N, the disk abundances are up to an order of magnitude higher than cometary for an adopted 30K rotational temperature; however, given the warmer (50K) HC$_3$N rotational temperature derived for MWC 480, we expect that the abundances calculated assuming a 30K temperature over-estimate the HC$_3$N/HCN ratio for the Herbig Ae disks at least.  Restricting the comparison to the 50--70K values, the HC$_3$N abundances are quite close to cometary.  

Figure \ref{fig_ratios}c shows the range of CH$_3$CN/HC$_3$N ratios measured in a sample of 16 low-mass protostellar envelopes \citep{Bergner2017}.  HCN column densities towards these sources are not available, however the CH$_3$CN/HC$_3$N ratio still provides a useful proxy for the relative efficiency of different complex nitrile chemistries.  The ratios across the disk sample are mostly consistent with the values measured in protostellar envelopes, with the exception of V4046 Sgr which is somewhat enhanced in CH$_3$CN/HC$_3$N compared to the other disks and protostars. 

Based on this comparison, we see that the gas-phase nitrile abundances relative to other N-bearing molecules are consistent across various physical environments: disk molecular layers, protostellar envelopes, and the midplane of the solar nebula.  We note that with these observations alone we cannot directly compare the comet- and planet-forming material in our sample with that of the protosolar nebula, as this would require extrapolations (i) from the molecular layer down to the midplane, and (ii) from gas-phase to ice abundances.  Nonetheless, the consistency of nitrile abundances across a wide range of physical conditions demonstrates a robust nitrile chemistry with similar outcomes in different environments.  CH$_3$CN abundance ratios (and upper limits) in particular appear to be especially regular both across the disk sample and in comparison with comets and protostars.  Complex nitrile species should therefore be reliably produced in a variety of different star- and planet-forming environments.

While the abundances of N-bearing molecules appear internally consistent across a range of physical environments, there is evidence that the ratio of N- to O-bearing COMs in disks is distinct compared to other environments.  In both comets and protostellar envelopes the CH$_3$CN/CH$_3$OH ratio is typically on the order of a few percent \citep{Mumma2011, Bergner2017}.  By contrast, in the one disk where CH$_3$OH has been detected (TW Hya), the column density ratio of CH$_3$CN/CH$_3$OH is about unity \citep{Walsh2016, Loomis2017}, indicative of an oxygen-poor chemistry.  Similarly, our observations covered a number of CH$_3$OH transitions in the 5--4 ladder, with no CH$_3$OH detections despite the strong nitrile emission.  This suggests that the under-abundance of gas-phase O- vs. N-bearing COMs is systematic in disks.  A nitrogen-rich, oxygen-poor chemistry is qualitatively consistent with an oxygen-starved environment due to e.g. the depletion of H$_2$O and CO from the gas phase \citep{Du2015}.  Such a scenario would indicate a predominantly gas-phase formation pathway for CH$_3$CN in disks.  Another possible factor is if the photodesorption efficiency of intact CH$_3$CN is high compared to CH$_3$OH, which has been shown to photodesorb mainly as fragments \citep{Bertin2016, Cruz2016}.  Since photodesorption from grains is thought to be of primary importance in disks, compared to mainly thermal desorption in protostars and comets, this could also contribute to the observed discrepancy in CH$_3$CN/CH$_3$OH across circumstellar environments.  Further exploration of the nitrile formation chemistry in disks using astrochemical models is needed to resolve the origin of this unique chemistry.

\section{Conclusions}
Based on ALMA observations of the complex nitrile species CH$_3$CN and HC$_3$N towards six protoplanetary disks, we conclude the following:

\begin{enumerate}
\item Complex nitrile molecules are commonly observed in protoplanetary disks, with five of six disks detected in HC$_3$N and three of six disks detected in CH$_3$CN.
\item Rotational temperatures derived for sources with multiple line detections are consistent with emission from the temperate molecular layer of the disk.  V4046 Sgr exhibits cool (29 $\pm$ 2K) CH$_3$CN emission consistent with the temperature measured in TW Hya \citep{Loomis2017}.  CH$_3$CN and HC$_3$N in MWC 480 are both characterized by warmer emission, with rotational temperatures of 73 $\pm$ 23K and 49 $\pm$ 6K respectively.  The increased radiation field around Herbig Ae disks compared to T Tauri disks may be responsible for this difference.
\item Parametric models of the CH$_3$CN, HC$_3$N, and H$^{13}$CN abundances in MWC 480 and V4046 Sgr are used to fit the observed emission and constrain radial abundance profiles.  Within 100AU, CH$_3$CN/HCN abundances are on the order of a few percent and HC$_3$N/HCN abundances on the order of tens of percent.
\item Across the disk sample we observe a tentative correlation of CH$_3$CN with H$^{13}$CN emission; if confirmed by further detections, the formation chemistry of CH$_3$CN should be re-visited to explain this relationship.  We see evidence for a possible anti-correlation in the spatial distributions of HC$_3$N and its pre-cursor C$_2$H; if confirmed in lower-J HC$_3$N transitions this would seem to rule out the current proposed HC$_3$N formation path.
\item We use the heterogenous physical properties of our disk sample to explore whether the UV field, disk age, or presence of an inner dust cavity impact the nitrile chemistry.  We observe no strong trends relating these environmental properties to the nitrile emission strength.  We emphasize the need for observations of lower-energy HC$_3$N lines to help constrain any relationships with disk physical properties.
\item Disk-averaged CH$_3$CN and HC$_3$N abundances relative to other N-bearing molecules are compared to values measured in solar system comets and protostellar envelopes and found to be consistent across these different of environments, although the HC$_3$N/HCN uncertainties are large due to sensitivity to the adopted rotational temperature.  These molecules appear to be reliably produced under a wide variety of physical conditions, demonstrating a robust nitrogen chemistry with similar outcomes in different environments.
\item Our results are suggestive of a disk chemistry systematically rich in N-bearing relative to O-bearing COMs when compared to other circumstellar environments.  The origin of this unique chemistry observed in disks compared to other stages of star and planet formation remains to be resolved.
\end{enumerate}

\acknowledgments 
This paper makes use of ALMA data, project codes: 2013.1.00226 and 2013.1.01070.S. ALMA is a partnership of ESO (representing its member states), NSF (USA), and NINS (Japan), together with NRC (Canada) and NSC and ASIAA (Taiwan), in cooperation with the Republic of Chile. The Joint ALMA Observatory is operated by ESO, AUI/NRAO, and NAOJ. The National Radio Astronomy Observatory is a facility of the National Science Foundation operated under cooperative agreement by Associated Universities, Inc. 

This manuscript benefited greatly from discussions with Ted Bergin and Romane Le Gal.  J.B.B acknowledges funding from the National Science Foundation Graduate Research Fellowship under Grant DGE1144152.  V.G.G. acknowledges support from the National Aeronautics and Space Administration under grant No. 15XRP15\_20140 issued through the Exoplanets Research Progam.  K.I.\"O. acknowledges funding from the Simons Collaboration on the Origins of Life (SCOL), the Alfred P. Sloan Foundation, and the David and Lucile Packard Foundation.  RAL acknowledges funding from NRAO Student Observing Support.

\software{
{\fontfamily{qcr}\selectfont NumPy} \citep{VanDerWalt2011},
{\fontfamily{qcr}\selectfont Matplotlib} \citep{Hunter2007},
{\fontfamily{qcr}\selectfont Astropy} \citep{Astropy2013}, 
{\fontfamily{qcr}\selectfont emcee} \citep{Foreman-Mackey2013},
{\fontfamily{qcr}\selectfont RADMC-3D} \citep{Dullemond2012},
{\fontfamily{qcr}\selectfont scikit-image} \citep{VanDerWalt2014},
{\fontfamily{qcr}\selectfont vis\_sample} \citep{Loomis2017a}
}

\clearpage
\appendix

\section{Channel maps}
\label{app_chanmaps}
\begin{figure*}[h!]
	\includegraphics[width=\linewidth]{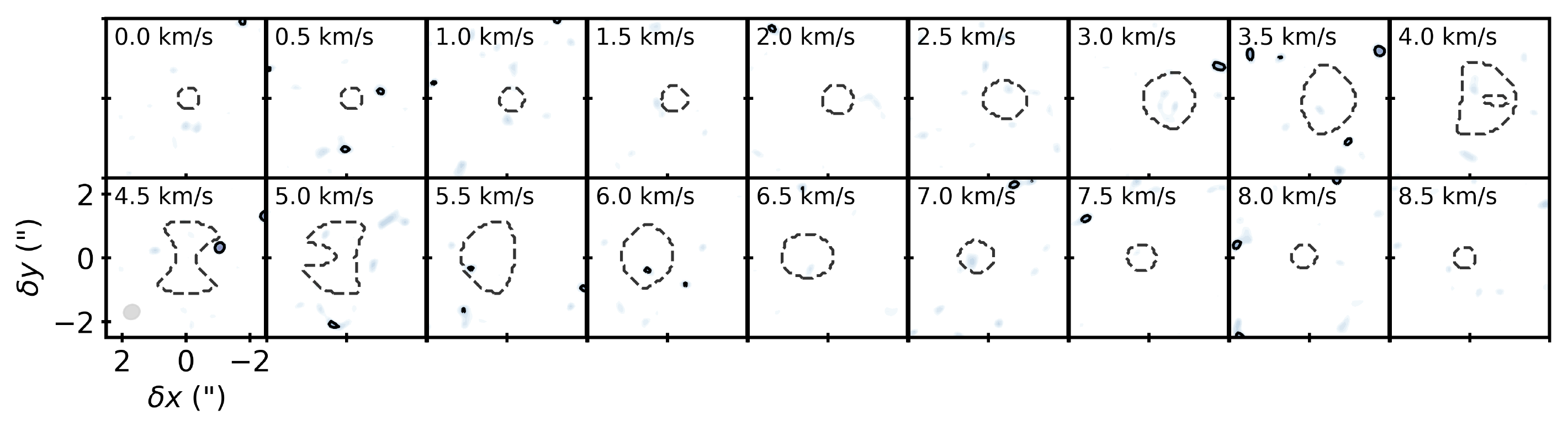}
	\caption{Same as Figure \ref{chan_hc3n_v4046sgr} but for CH$_3$CN 14$_0$--13$_0$ in AS 209.}
\label{chan_ch3cn_as209}
\end{figure*}

\begin{figure*}[h!]
	\includegraphics[width=\linewidth]{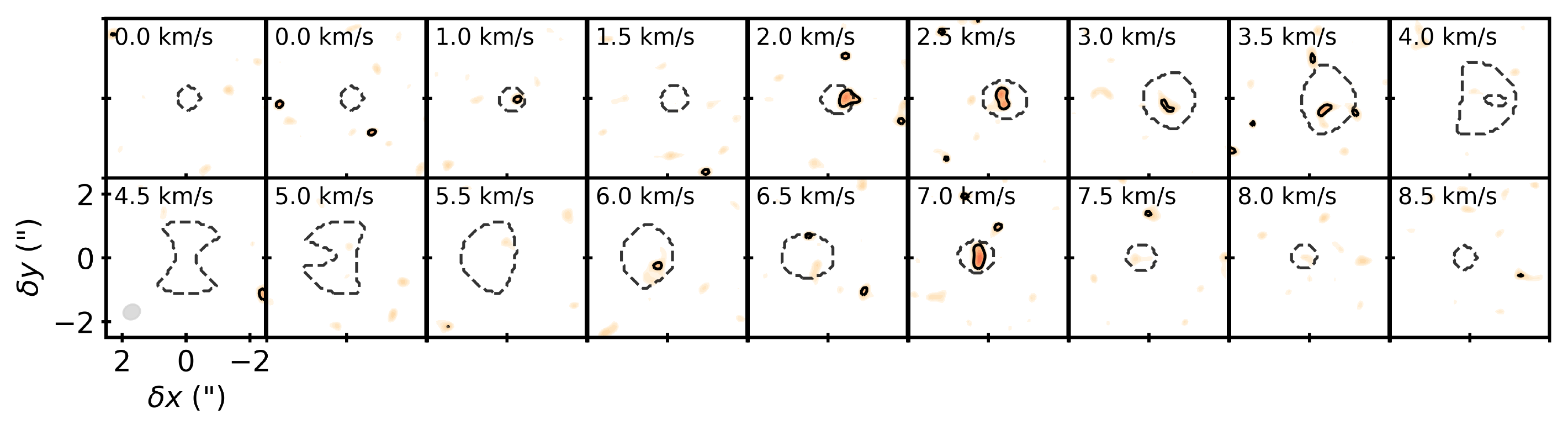}
	\caption{Same as Figure \ref{chan_hc3n_v4046sgr} but for HC$_3$N 27--26 in AS 209.} 
\label{chan_hc3n_as209}
\end{figure*}

\begin{figure*}[h!]
	\includegraphics[width=\linewidth]{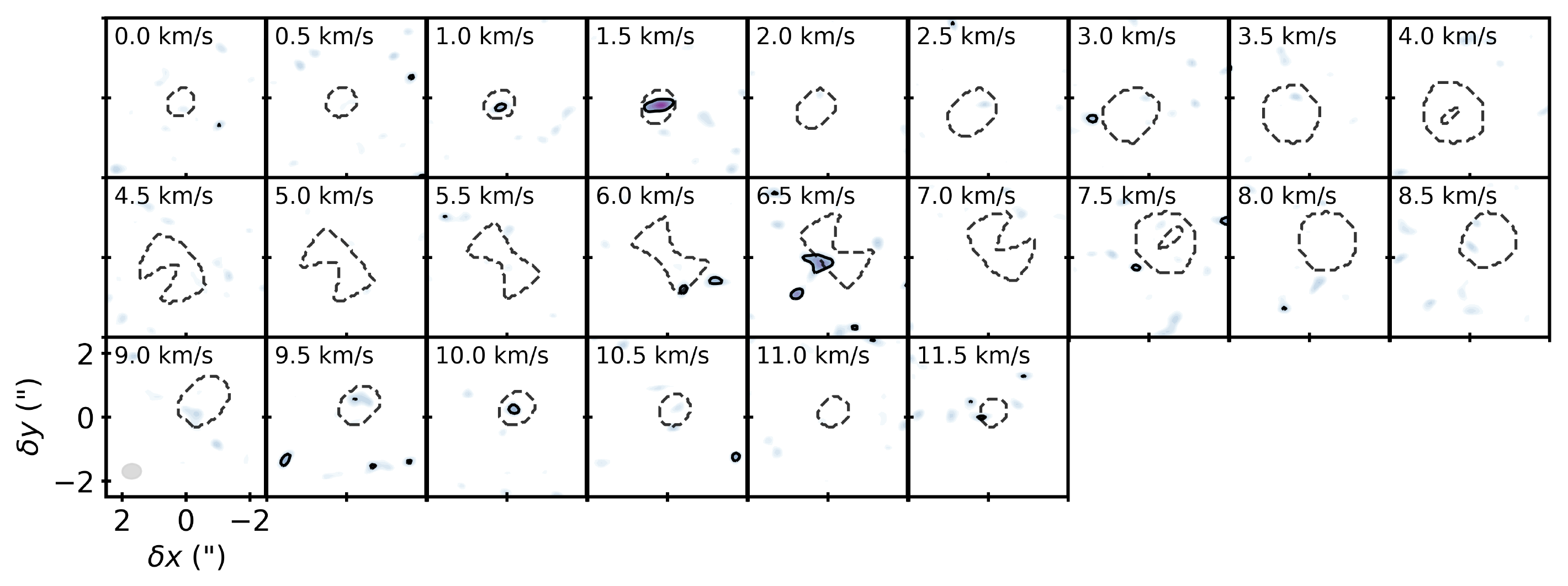}
	\caption{Same as Figure \ref{chan_hc3n_v4046sgr} but for CH$_3$CN 14$_0$--13$_0$ in HD 163296.}
\label{chan_ch3cn_hd163296}
\end{figure*}

\begin{figure*}[h!]
	\includegraphics[width=\linewidth]{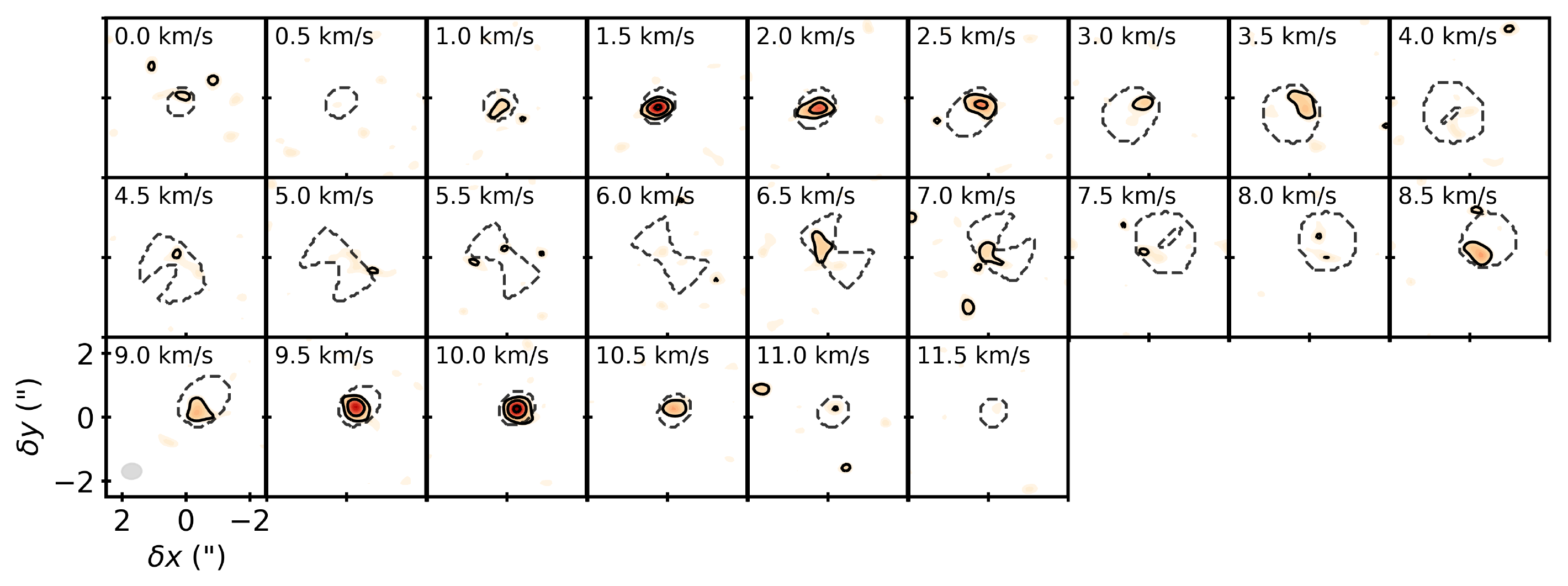}
	\caption{Same as Figure \ref{chan_hc3n_v4046sgr} but for HD 163296 HC$_3$N 27--26.}
\label{chan_hc3n_hd163296}
\end{figure*}

\begin{figure*}[h!]
	\includegraphics[width=\linewidth]{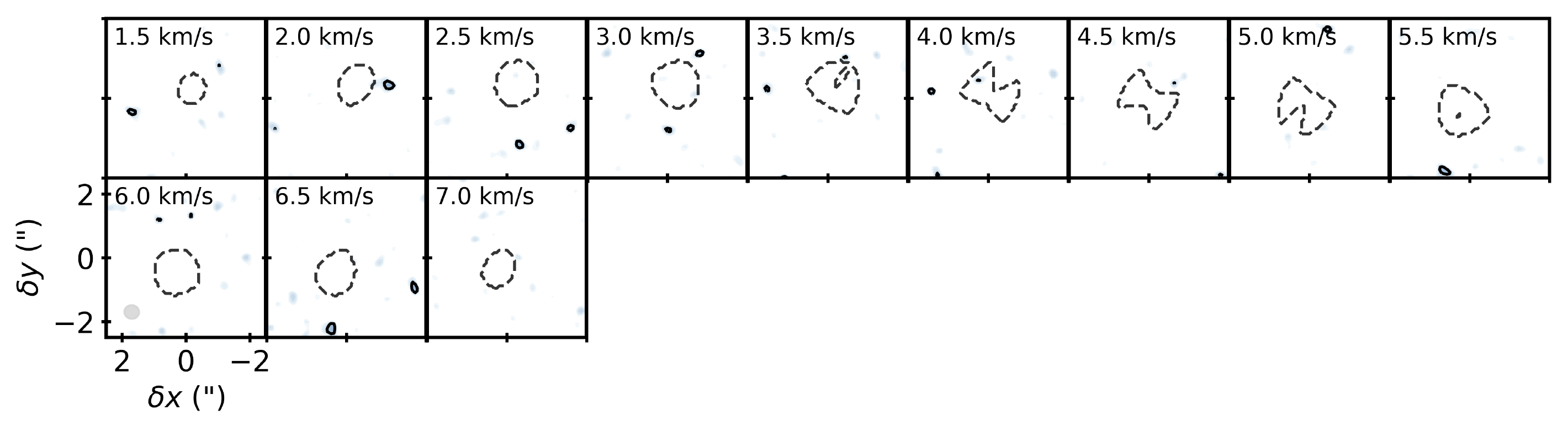}
	\caption{Same as Figure \ref{chan_hc3n_v4046sgr} but for CH$_3$CN 14$_0$--13$_0$ in IM Lup.}
\label{chan_ch3cn_imlup}
\end{figure*}

\begin{figure*}[h!]
	\includegraphics[width=\linewidth]{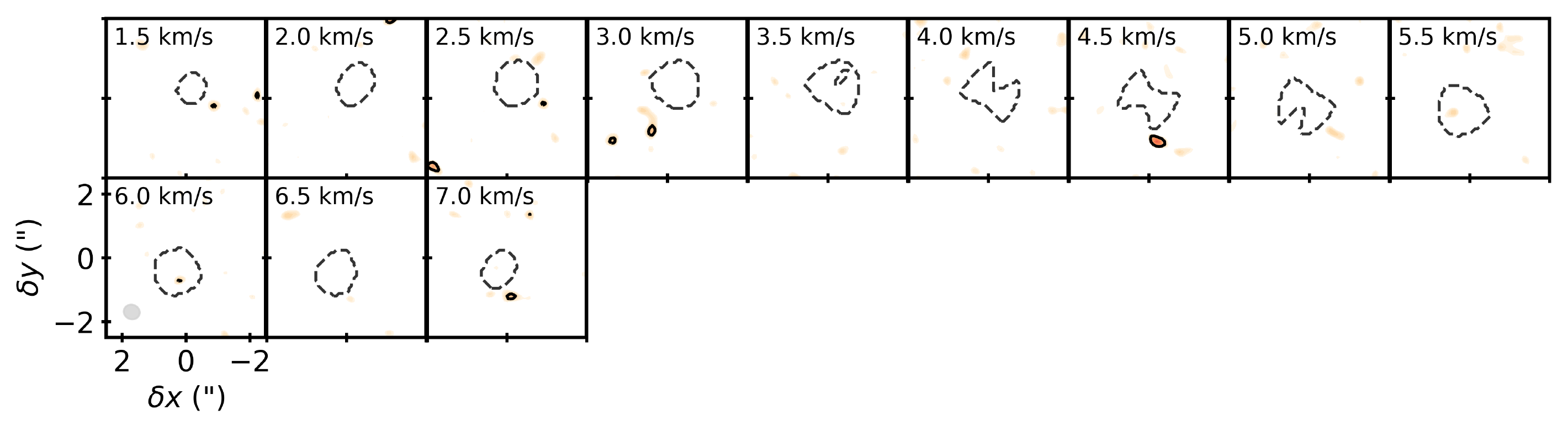}
	\caption{Same as Figure \ref{chan_hc3n_v4046sgr} but for HC$_3$N 27--26 in IM Lup.}
\label{chan_hc3n_imlup}
\end{figure*}

\begin{figure*}[h!]
	\includegraphics[width=\linewidth]{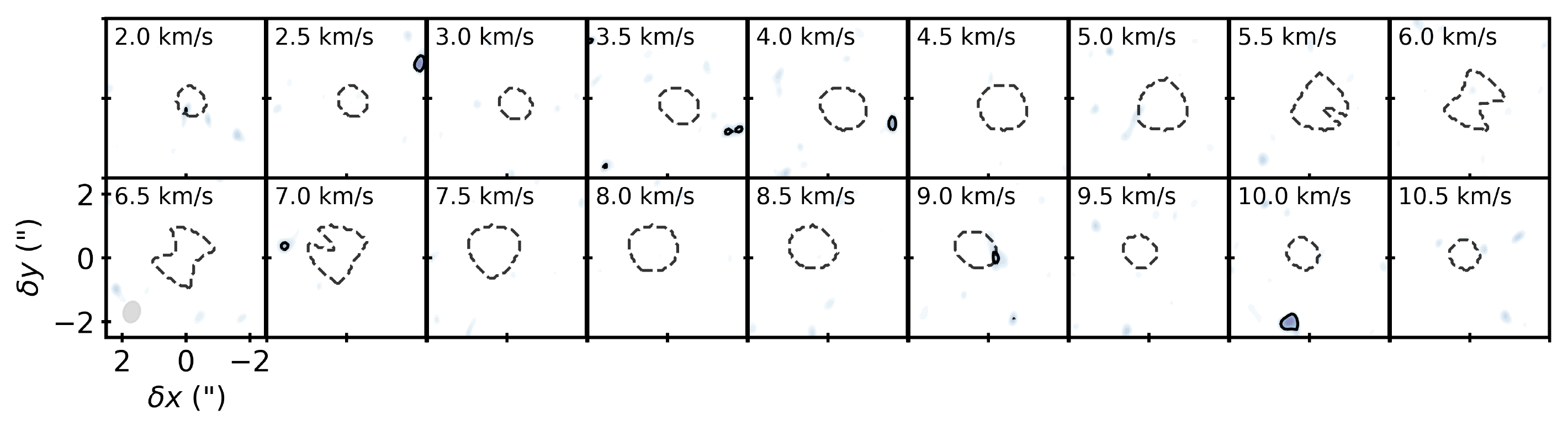}
	\caption{Same as Figure \ref{chan_hc3n_v4046sgr} but for CH$_3$CN 14$_0$--13$_0$ in LkCa 15.}
\label{chan_ch3cn_lkca15}
\end{figure*}

\begin{figure*}[h!]
	\includegraphics[width=\linewidth]{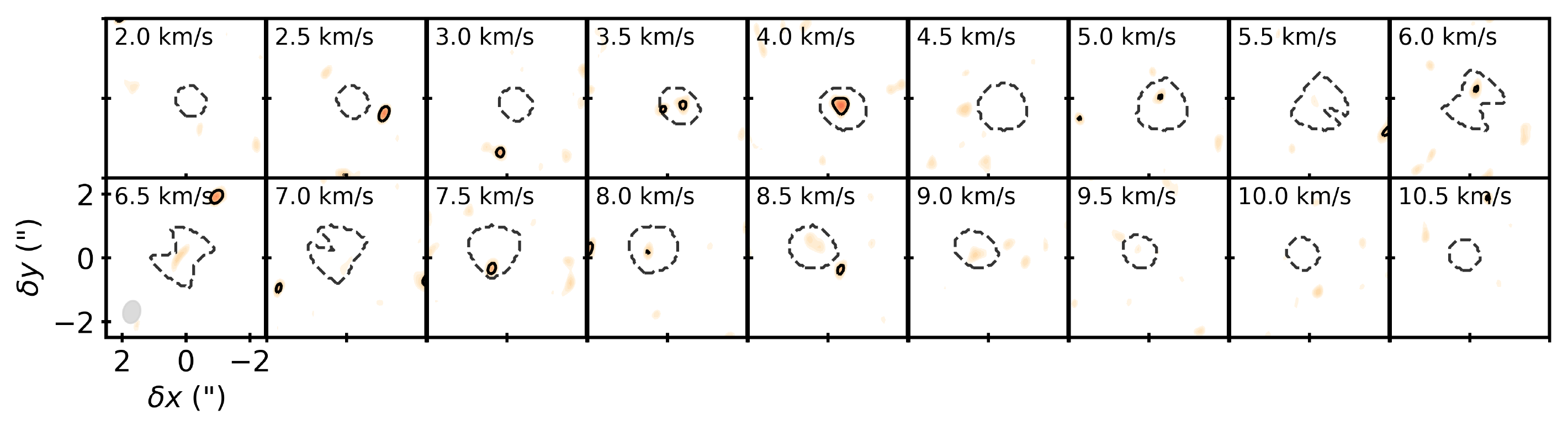}
	\caption{Same as Figure \ref{chan_hc3n_v4046sgr} but for HC$_3$N 27--26 in LkCa 15.}
\label{chan_hc3n_lkca15}
\end{figure*}

\begin{figure*}[h!]
	\includegraphics[width=\linewidth]{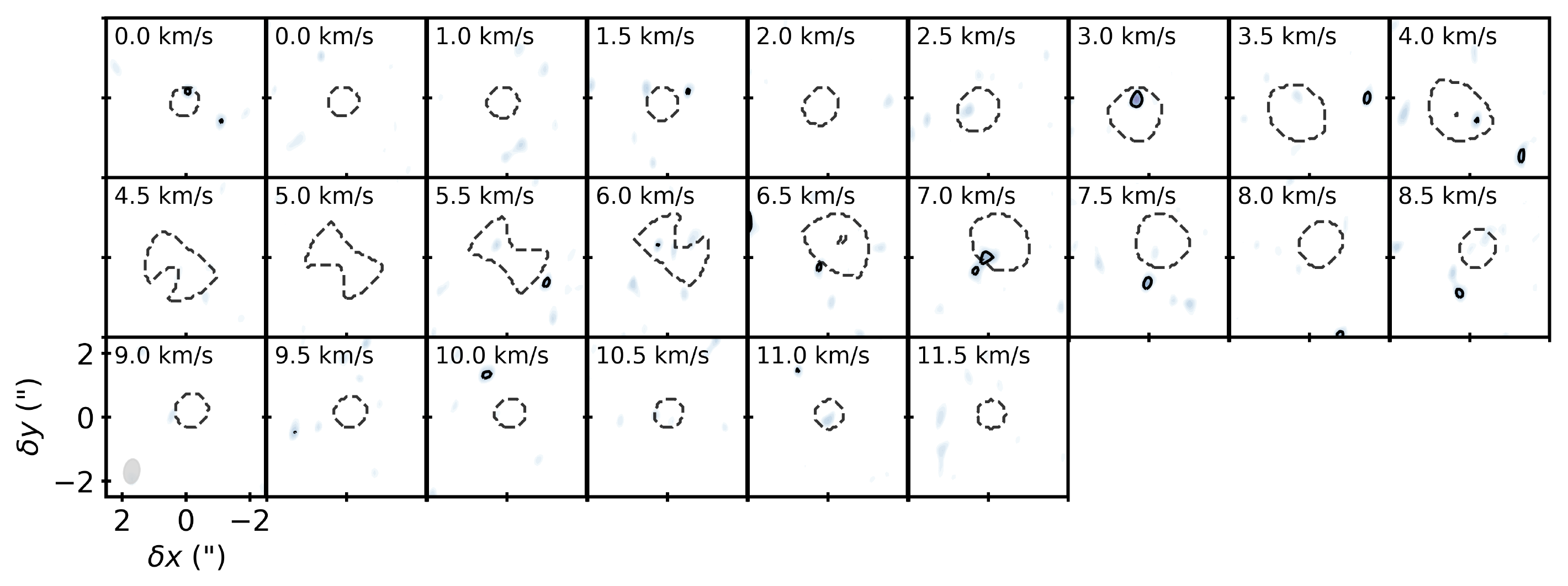}
	\caption{Same as Figure \ref{chan_hc3n_v4046sgr} but for CH$_3$CN 14$_0$--13$_0$ in MWC 480.}
\label{chan_ch3cn_mwc480}
\end{figure*}

\begin{figure*}[h!]
	\includegraphics[width=\linewidth]{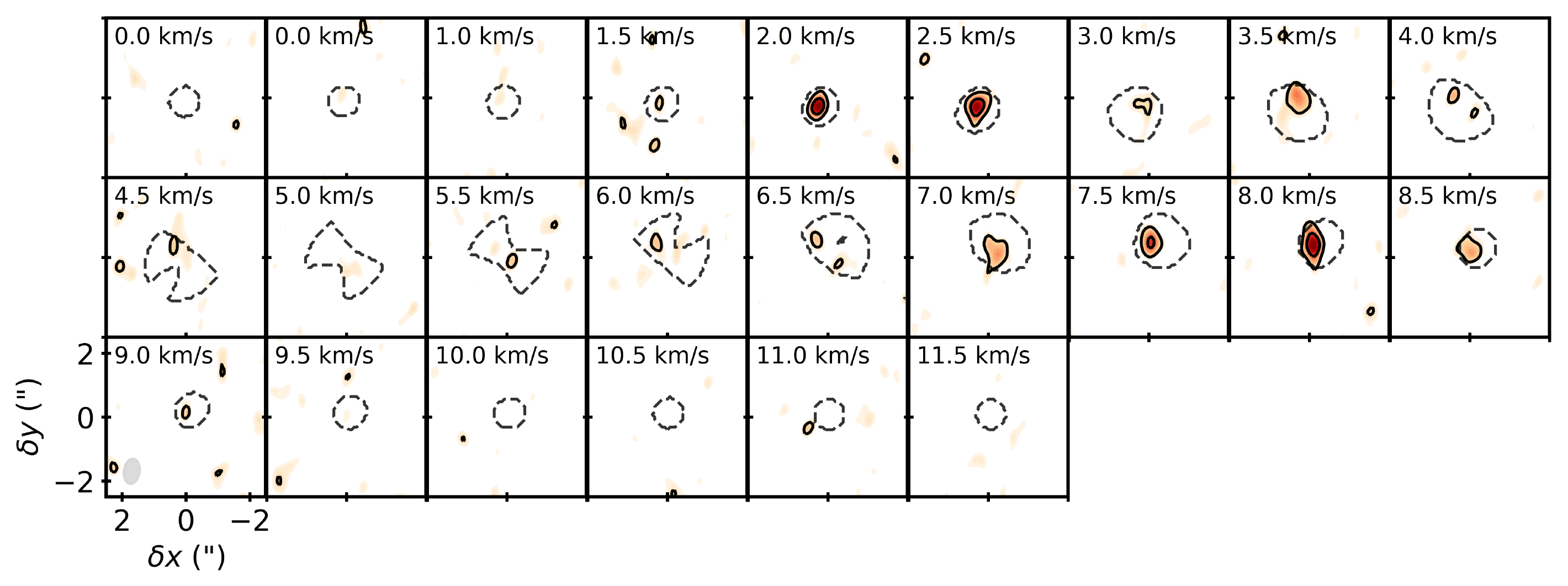}
	\caption{Same as Figure \ref{chan_hc3n_v4046sgr} but for HC$_3$N 27--26 in MWC 480.}
\label{chan_hc3n_mwc480}
\end{figure*}

\begin{figure*}[h!]
	\includegraphics[width=\linewidth]{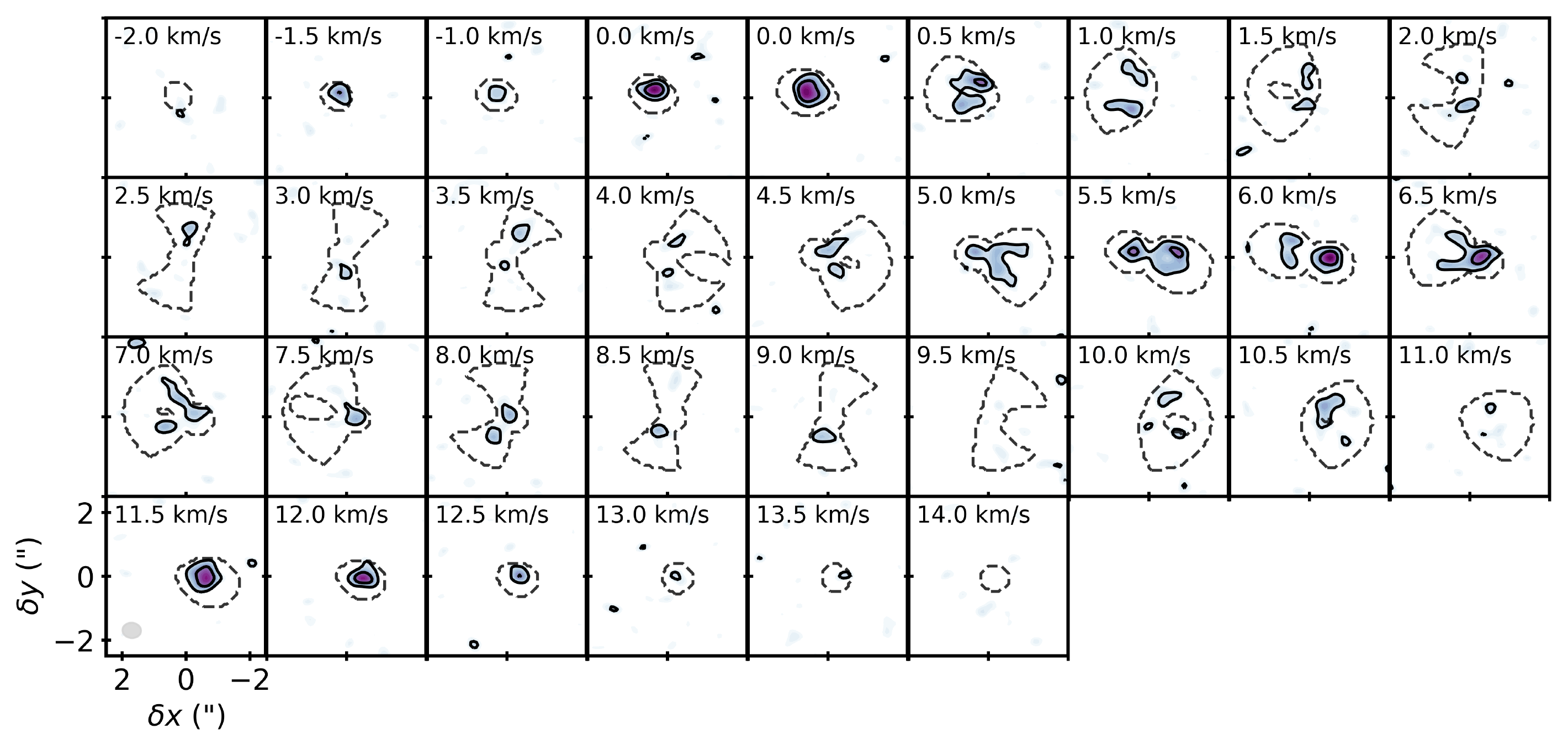}
	\caption{Same as Figure \ref{chan_hc3n_v4046sgr} but for CH$_3$CN 14$_0$--13$_0$ and 14$_1$--13$_1$ in V4046 Sgr.}
\label{chan_ch3cn_v4046sgr}
\end{figure*}

\begin{figure*}[h!]
	\includegraphics[width=\linewidth]{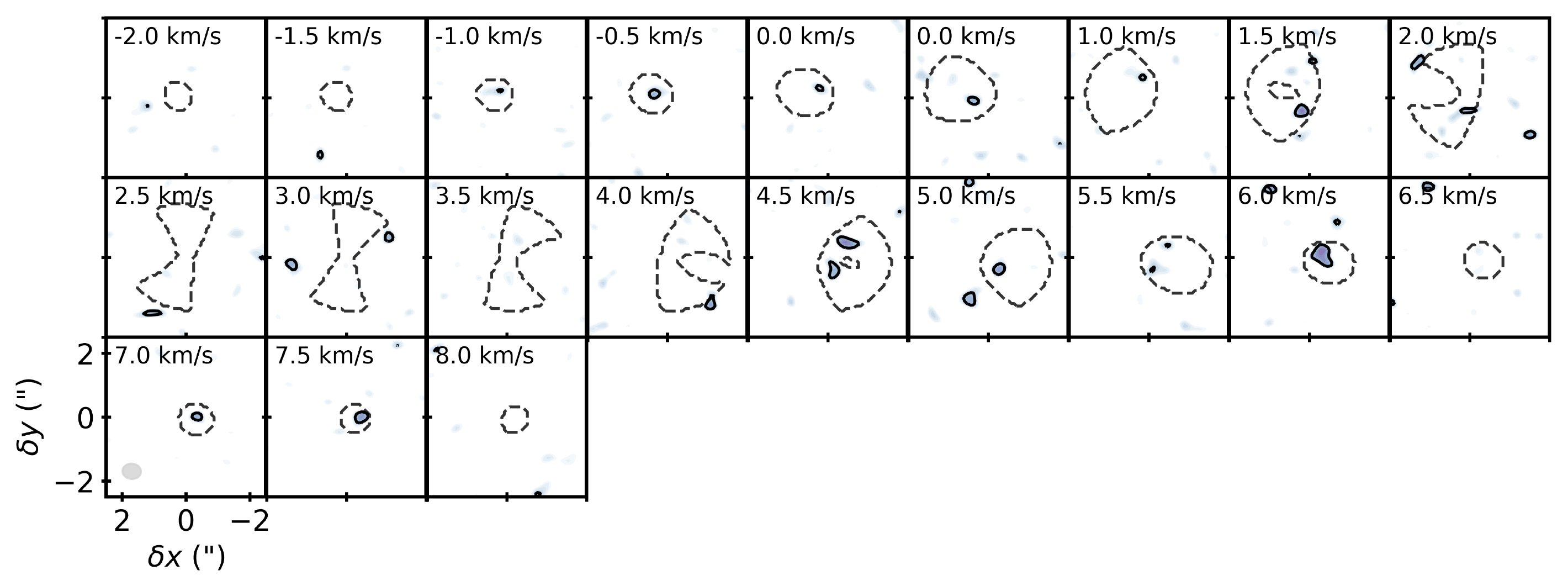}
	\caption{Same as Figure \ref{chan_hc3n_v4046sgr} but for CH$_3$CN 14$_2$--13$_2$ in V4046 Sgr.}
\label{chan_ch3cn4_v4046sgr}
\end{figure*}

\begin{figure*}[h!]
	\includegraphics[width=\linewidth]{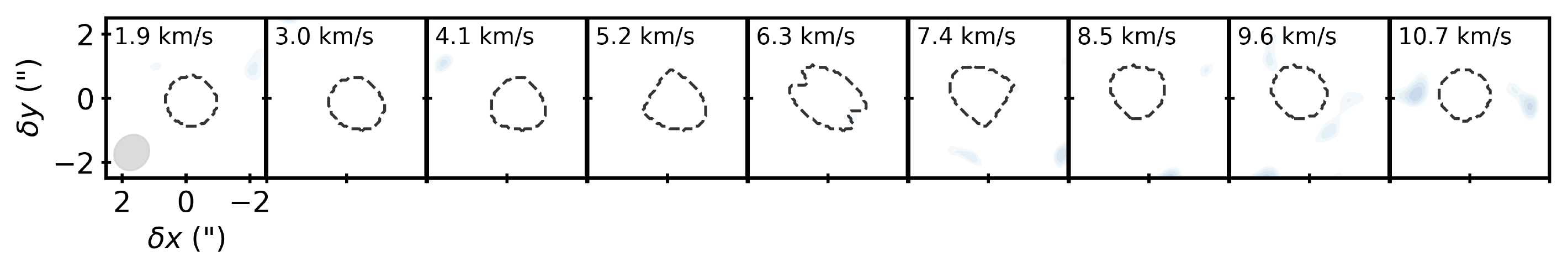}
	\caption{Same as Figure \ref{chan_hc3n_v4046sgr} but for CH$_3$CN 15$_0$--14$_0$ in LkCa 15.}
\label{chan_ch3cn2_lkca15}
\end{figure*}

\begin{figure*}[h!]
	\includegraphics[width=\linewidth]{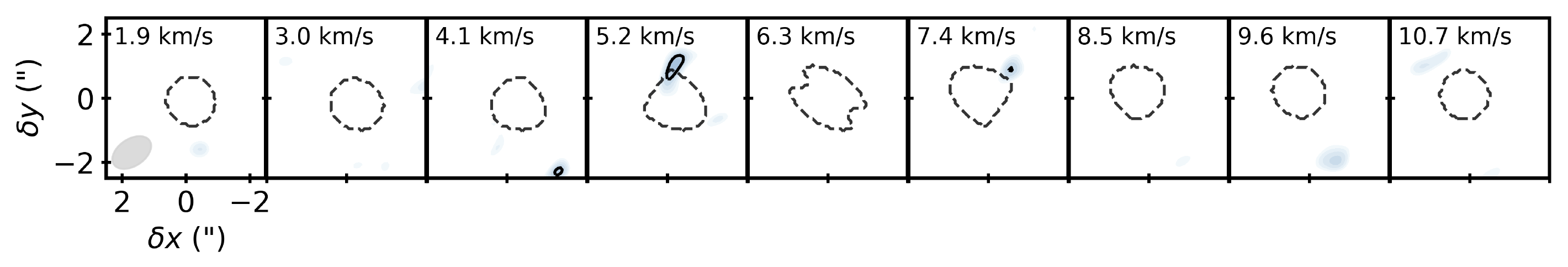}
	\caption{Same as Figure \ref{chan_hc3n_v4046sgr} but for CH$_3$CN 16$_0$--15$_0$ in LkCa 15.}
\label{chan_ch3cn3_lkca15}
\end{figure*}

\begin{figure*}[h!]
	\includegraphics[width=\linewidth]{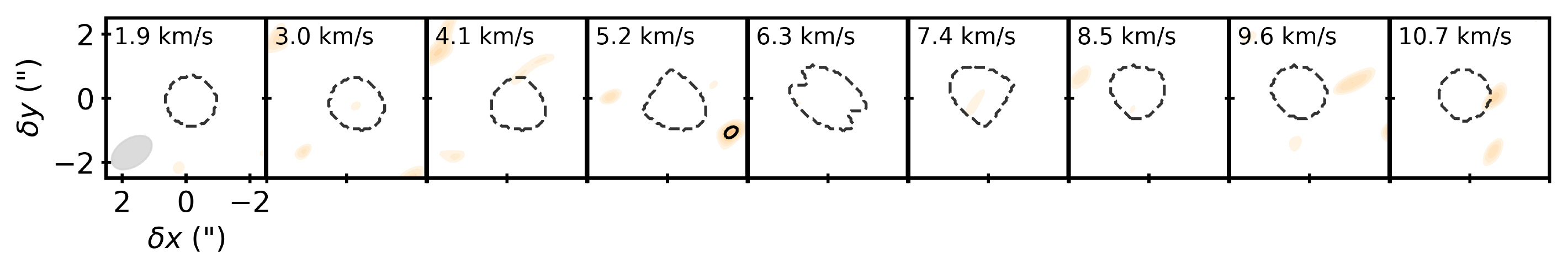}
	\caption{Same as Figure \ref{chan_hc3n_v4046sgr} but for HC$_3$N 31--30 in LkCa 15.}
\label{chan_hc3n1_lkca15}
\end{figure*}

\begin{figure*}[h!]
	\includegraphics[width=\linewidth]{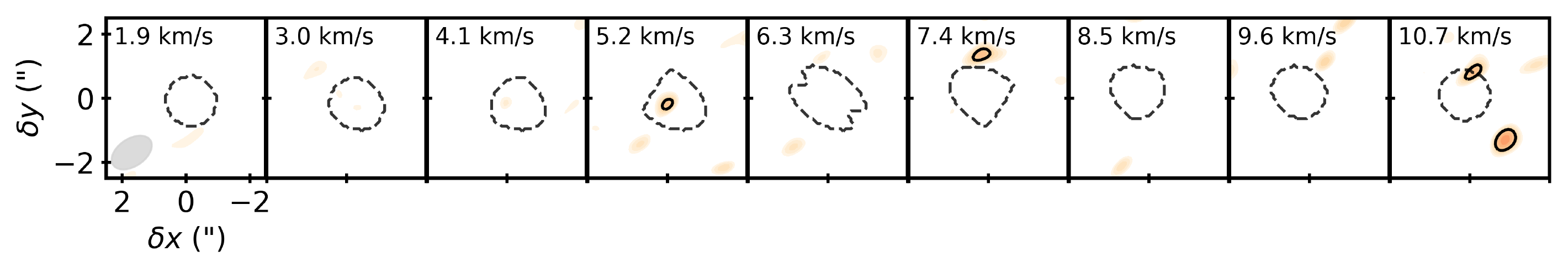}
	\caption{Same as Figure \ref{chan_hc3n_v4046sgr} but for HC$_3$N 32--31 in LkCa 15.}
\label{chan_hc3n2_lkca15}
\end{figure*}

\begin{figure*}[h!]
	\includegraphics[width=\linewidth]{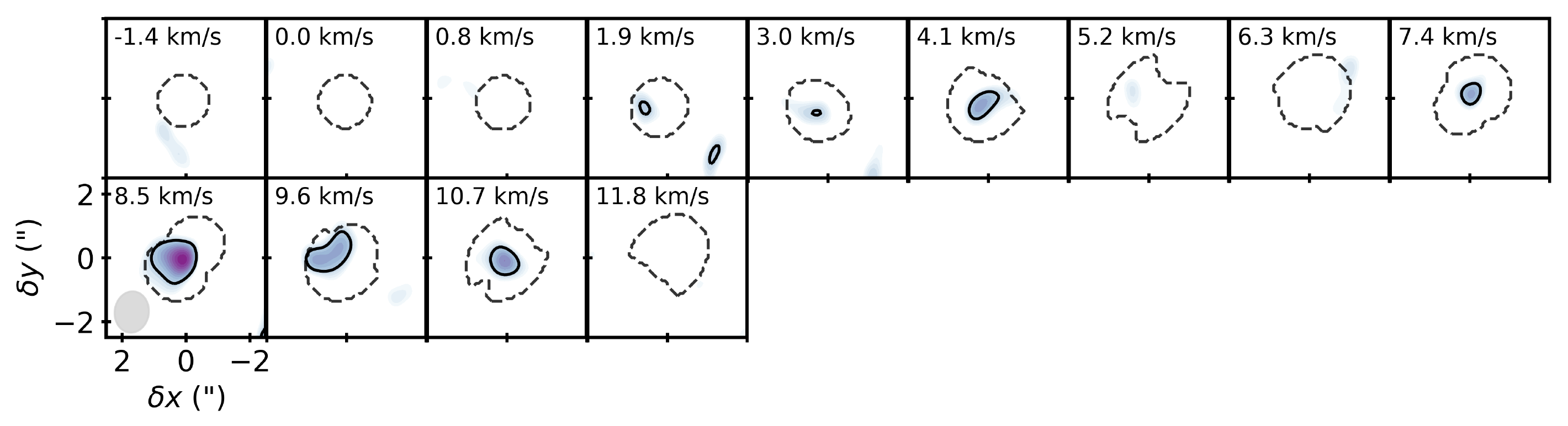}
	\caption{Same as Figure \ref{chan_hc3n_v4046sgr} but for CH$_3$CN 15$_0$--14$_0$ and 15$_1$--14$_1$ in MWC 480.}
\label{chan_ch3cn2_mwc480}
\end{figure*}

\begin{figure*}[h!]
	\includegraphics[width=\linewidth]{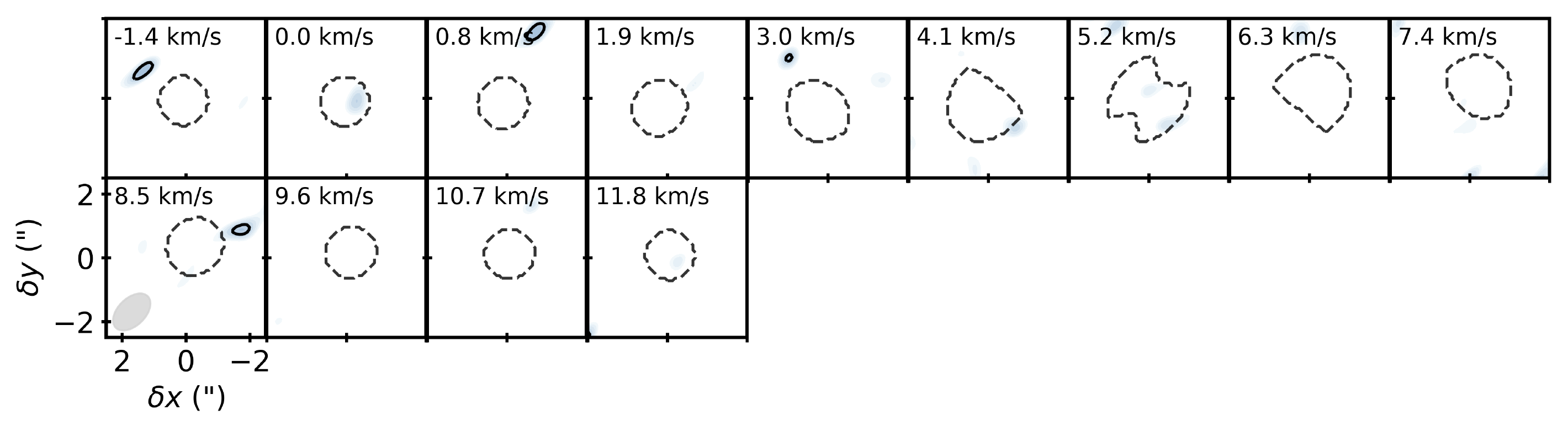}
	\caption{Same as Figure \ref{chan_hc3n_v4046sgr} but for CH$_3$CN 16$_0$--15$_0$ in MWC 480.}
\label{chan_ch3cn3_mwc480}
\end{figure*}

\begin{figure*}[h!]
	\includegraphics[width=\linewidth]{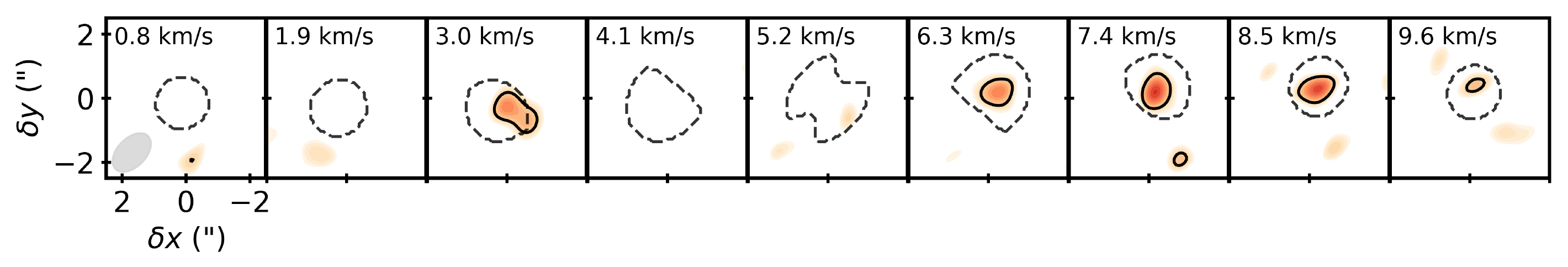}
	\caption{Same as Figure \ref{chan_hc3n_v4046sgr} but for HC$_3$N 31--30 in MWC 480.}
\label{chan_hc3n1_mwc480}
\end{figure*}

\begin{figure*}[h!]
	\includegraphics[width=\linewidth]{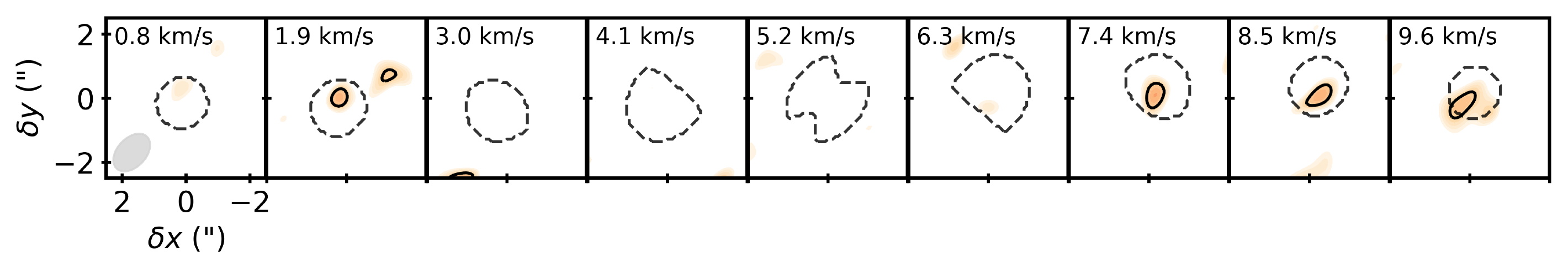}
	\caption{Same as Figure \ref{chan_hc3n_v4046sgr} but for HC$_3$N 32--31 in MWC 480.}
\label{chan_hc3n2_mwc480}
\end{figure*}

Channel maps for CH$_3$CN 14$_0$-13$_0$ and HC$_3$N 27-26 in all sources are shown in Figures \ref{chan_ch3cn_as209}-\ref{chan_ch3cn_v4046sgr}.  Channel maps for additional observed lines in V4046 Sgr, LkCa 15, and  MWC 480 are shown in Figures \ref{chan_ch3cn4_v4046sgr}-\ref{chan_hc3n2_mwc480}.

\clearpage
\section{Abundance models}
\label{app_models}
\begin{figure*}[h!]
	\includegraphics[width=\linewidth]{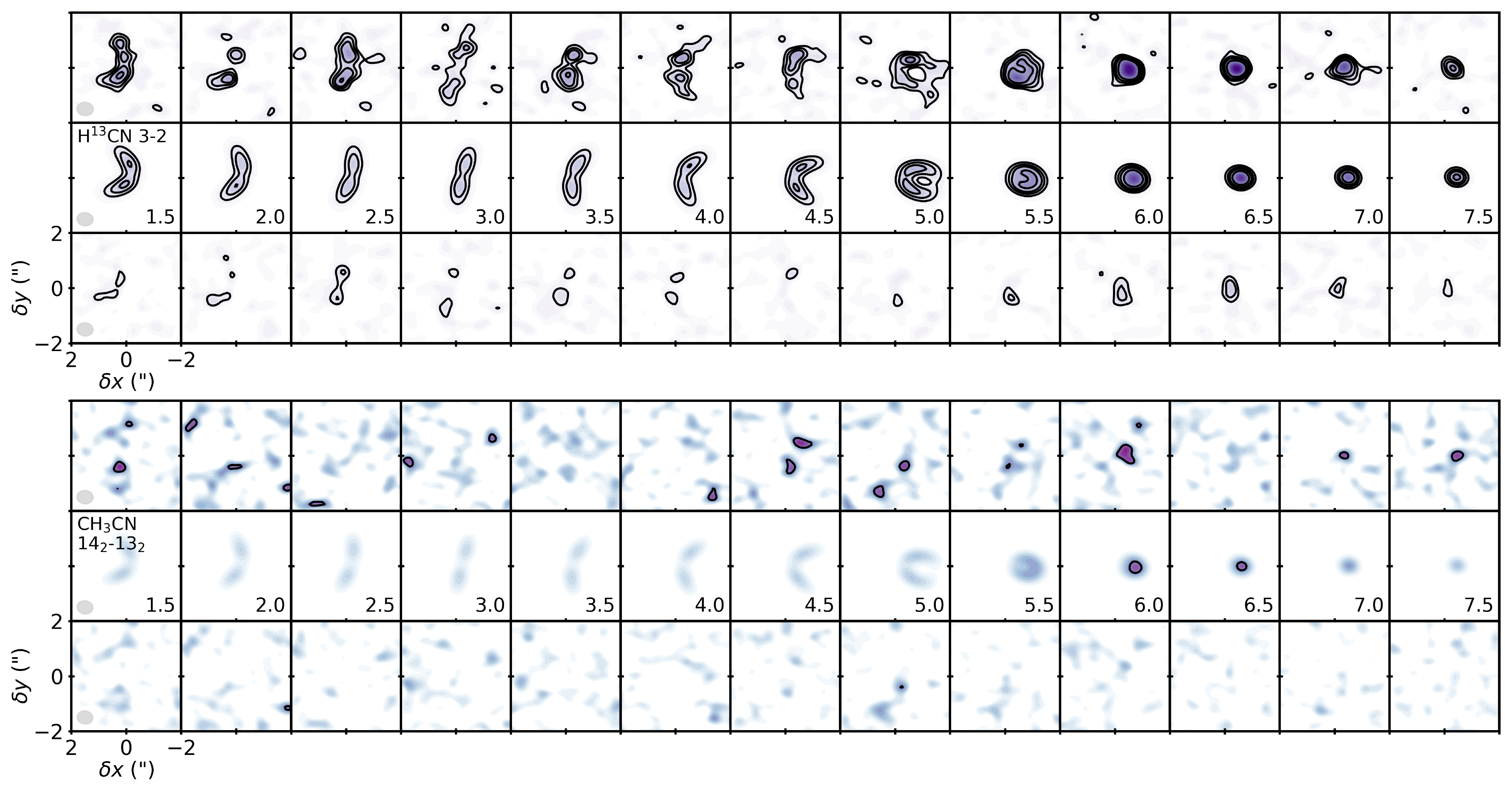}
	\caption{Observations (top), model (middle), and residuals (bottom) for H$^{13}$CN 3--2 and CH$_3$CN 14$_2$--13$_2$ in V4046 Sgr.  Contour levels indicate 3, 5, 7, 10, and 20$\times$ rms.}
\label{fig_vmodel_extras}
\end{figure*}

\begin{figure*}[h!]
	\includegraphics[width=0.7\linewidth]{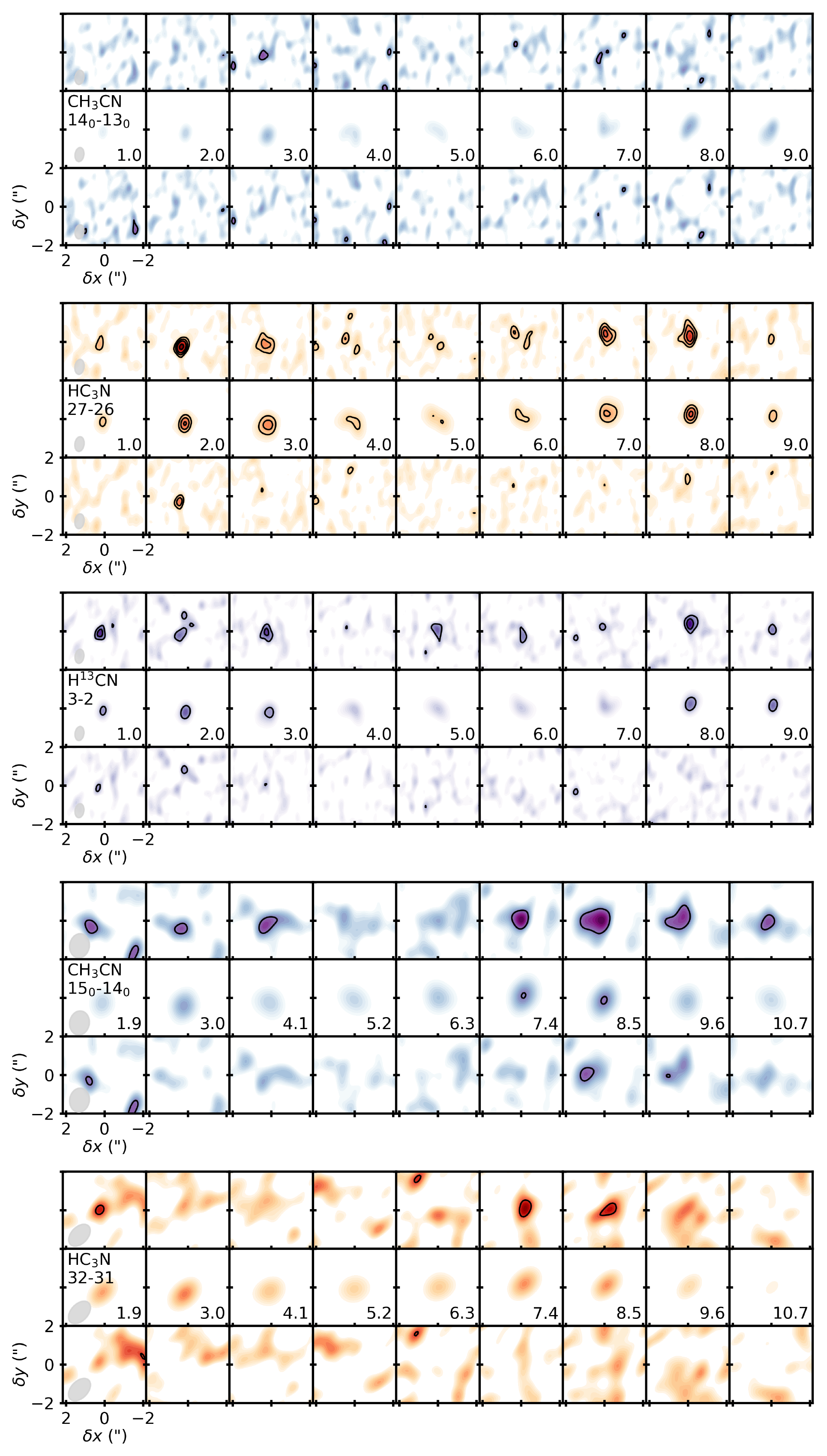}
	\caption{Observations (top), model (middle), and residuals (bottom) for CH$_3$CN 14$_0$--13$_0$, HC$_3$N 27--26, H$^{13}$CN 3--2, CH$_3$CN 15$_0$--14$_0$, and HC$_3$N 31--30 in MWC 480.  Contour levels indicate 3, 5, 7, and 10$\times$ rms.}
\label{fig_mmodel_extras}
\end{figure*}

Modeling results for H$^{13}$CN and CH$_3$CN 14$_2$--13$_2$ in V4046 Sgr are shown in Figure \ref{fig_vmodel_extras}.  Figure \ref{fig_mmodel_extras} shows modeling results for MWC 480 CH$_3$CN 14$_0$--13$_0$, HC$_3$N 27--26, H$^{13}$CN 3--2, CH$_3$CN 15$_0$--14$_0$, and HC$_3$N 31--30.

\end{document}